\documentclass[epj]{svjour}
\usepackage{epsfig}
\begin{document}

\title{Analysis of recent {\boldmath$\eta$}  photoproduction
data}
\author{A.~Sibirtsev\inst{1,2,3,4}, J.~Haidenbauer\inst{3,5},
S.~Krewald\inst{3,5} and
U.-G.~Mei{\ss}ner\inst{1,3,5}
}                     


\institute{
Helmholtz-Institut f\"ur Strahlen- und Kernphysik (Theorie)
und Bethe Center for Theoretical Physics,
Universit\"at Bonn, D-53115 Bonn, Germany  \and
Excited Baryon Analysis Center (EBAC), Thomas Jefferson National
Accelerator
Facility, Newport News, Virginia 23606, USA
\and Institut f\"ur Kernphysik and J\"ulich Center for Hadron Physics,
Forschungszentrum J\"ulich, D-52425 J\"ulich, Germany
\and Institute of High Energy Physics and Center for Theoretical Studies,
Chinese Academy of Sciences,
Beijing 100049, China
\and Institute for Advanced Simulation,
Forschungszentrum J\"ulich, D-52425 J\"ulich, Germany
}

\date{Received: date / Revised version: date}

\abstract{Recent data on $\eta$-meson photoproduction off
a proton target in the energy range $2{\le}\sqrt{s}{\le}3$~GeV
are analyzed with regard to their overall consistency. 
Results from the ELSA and CLAS measurements are 
compared with predictions of a Regge model whose reaction amplitude 
was fixed via a global fit to pre-2000 measurements
of differential cross sections and polarization observables
for $\gamma{p}{\to}\eta p$ 
at higher energies. We find that all recent experimental results on 
differential cross sections for $\eta$-meson photoproduction 
are in good agreement with each other, except for the CLAS data from 2009.
However, the latter can be made consistent with the other data at the 
expense of introducing an energy-dependent renormalization factor.
We point out that there indications in the data for a possible 
excitation of baryon resonances with masses around 2.1 and 2.4 GeV.
} 

\authorrunning{A. Sibirtsev {\it et al.}}

\maketitle

\section{Introduction}
Recently data on $\gamma{p}{\to}\eta p$ differential cross sections
were published by the ELSA~\cite{Crede09} and CLAS
Collaborations~\cite{Williams09}. It turned out that there is a
substantial disagreement between the two measurements. In fact, 
as indicated in Refs.~\cite{Crede09,Williams09}, the experimental
results were cross checked applying various methods but no plausible 
reason was found for the discrepancy.
The status of the data remains the same - they are inconsistent.

This situation is rather unsatisfactory. The data from both
measurements are given with high statistical and systematic accuracy and
with small increments in energy and angle. The ELSA measurement covers 
invariant collision energies up to $\sqrt{s}\simeq$2.37 GeV, while the
highest energy considered in the CLAS experiment is $\sqrt{s}=$2.795 GeV. 
Therefore, these new data
would allow to study possible excitation of high-mass baryons. The
$\eta N$ state couples to baryons with isospin $1/2$ and it might be that
$\eta$-meson photoproduction is more selective to resonance excitations
than pion photoproduction or $\pi N$ scattering. A typical example
is the $P_{11}(1710)$ resonance that was not seen in many analyses of
elastic $\pi N$ scattering but appears to become very important once one
includes the $\pi N\to \eta N$ channel~\cite{Ceci06} and works in a
coupled-channel formalism, see also the combined analysis of  
$\pi N$ scattering and photoproduction data in \cite{Sarantsev:2007bk}.

For reactions where a large data base exists one can formulate rejection
criteria for inconsistent data sets solely based on statistical arguments.
This was done, for example, in partial wave analyses of the $NN$ 
\cite{Bergervoet} and $\bar NN$ systems \cite{Timmermans}.
Unfortunately, for the reaction $\gamma{p}{\to}\eta p$ one is far away from
such a situation. Here, one simply would have to exclude one or the other
measurement from the global data analysis without clear criterion. 
Apparently, such a procedure is not very appealing and, moreover, 
disregards the experimental progress in studying the spectrum of excited 
baryons. 
In any case, as an alternative, one should examine whether floating the 
normalization of some experimental results is a possible way to resolve 
inconsistencies between the data. Even in the analysis of reactions with
a large data base, like $NN$ or $\pi N$, it is standard practice to allow
for 
some 
variation in the normalization of data sets
\cite{Bergervoet,Arndt72,Arndt06}. 

A direct comparison of the different $\gamma p \to \eta p$ 
measurements is complicated because,
in general, data were taken at different energies and different angles in 
the ELSA and CLAS experiments. One possibility is to compare the observed 
differential cross sections with model predictions. Phenomenological
analyses 
along this line were presented in Refs.~\cite{Adelseck90,Mart07} in an
attempt to 
examine the consistency of $K\Lambda$ photoproduction data. It is clear 
that any chosen specific model might not be able to describe the data 
fully satisfactorily and, thus, will deviate necessarily from the data to 
some extent. 
However, one expects that any such deviation between the model calculation 
and different measurements would occur systematically and on a similar 
quantitative level if the experimental results are mutually consistent. 
In the present paper we adopt this idea~\cite{Adelseck90,Mart07}
and perform an analysis of the reaction $\gamma{p}{\to}\eta p$ in the 
same spirit. 
Specifically, we compare the results from the ELSA and CLAS 
measurements with predictions of a Regge model \cite{Sibirtsev11}
whose reaction amplitude was fixed via a global fit to pre-2000
measurements
of differential cross sections and polarizations for $\gamma{p}{\to}\eta
p$ 
at higher energies, i.e. at photon energies above 3 GeV ($\sqrt{s}>$2.55
GeV). 

Apart from analyzing the consistency of the data from different 
measurements we are also interested in the search for excited baryons 
with masses above 1.9 GeV.
Indeed, possible signals for the excitation of baryon resonances at
$\sqrt{s}\ge$1.9 GeV were discussed in the first CLAS
publication~\cite{Dugger02} on $\eta$-meson photoproduction. Furthermore,
the systematic study of single-pion photoproduction presented 
in~\cite{Sibirtsev08,Sibirtsev09} indicates also possible signals for 
resonances at invariant energies around 2 GeV.

There are few excited baryons with a mass above 1.9 GeV listed by
the Particle Data Group (PDG)~\cite{PDG} that possibly couple to the 
$\eta N$ channel. These resonances are the $S_{11}(2090)$, the
$P_{11}(2110)$ 
and the $G_{17}(2190)$. They were found in multichannel
analyses~\cite{Vrana00,Batinic95}. It is important to note that in
Ref.~\cite{Vrana00} the objective was to evaluate the resonance
parameters from available pion-nucleon partial wave analyses (PWA). 
The issue 
of extracting such partial wave amplitudes from the observables was not
considered in this paper. The evaluation procedure adopted in
Ref.~\cite{Batinic95} is based on a multi-resonance model fitted to
available partial waves obtained from $\pi N$ scattering data. Therefore,
the presently available PDG information~\cite{PDG} concerning excited
high-mass 
baryons coupled to the $\eta N$-channel is essentially based on PWA's of
pion-nucleon scattering data~\cite{Cutkosky79,Koch80,Hoehler83}. 

The paper is organized as follows. 
In Sect. 2 we describe in detail the reaction amplitudes employed in
the present investigation. 
An analysis of the experimental results from ELSA and CLAS is provided in 
Sects. 3 and 4, respectively. 
In Sect. 5 we review the current status of the data available for 
the $\gamma{p}{\to}\eta p$ differential cross section and we investigate
the consistency of those data in detail. 
The paper ends with a short summary. Quantitative results of the comparision
of the different measurements with the predictions of our Regge model 
are summarized in an Appendix.

\section{The reaction amplitude}

Similar to our previous analyses of charged and neutral pion
photoproduction~\cite{Sibirtsev08,Sibirtsev07} we use a gauge invariant
Regge model, which combines the Regge pole and cut amplitudes for $\rho$,
$\omega$ and $b_1$ exchanges.  At high energies the interactions before and
after the basic Regge pole exchange mechanisms are essentially elastic or
diffractive scattering described by Pomeron exchange. Such a scenario can
be related to the distorted wave approximation and provides a well defined
formulation for constructing Regge cut amplitudes.

We use the $t$-channel parity conserving helicity amplitudes $F_i$ ($i{=}
1, ..., 4$).  Here $F_1$ and $F_2$ are the natural and unnatural
spin-parity
$t$-channel amplitudes to all orders in $s$, respectively. $F_3$ and $F_4$
are
the natural and unnatural $t$-channel amplitudes to leading order in $s$. 

Each Regge pole helicity amplitude is parameterized by 
\begin{eqnarray}
F (s,t) =  \pi \beta(t)   \frac{1 {+} {\cal S} \exp[-i \pi
\alpha
(t)]}{\sin[\pi\alpha(t)] \,\,\Gamma[\alpha(t)]}
\left[\frac{s}{s_0}\right]^{\alpha(t)-1},
\label{rpropa}
\end{eqnarray}
where $s$ is the invariant collision energy squared, $t$ is the 
four-momentum transfer squared and $s_0$=1~GeV$^2$ is a scaling parameter.
Furthermore, $\beta (t)$ is a residue function, ${\cal S}$ is the signature
factor and $\alpha(t)$ is the Regge trajectory. 

The structure of the vertex function $\beta(t)$ of Eq.~(\ref{rpropa})
is defined by the quantum numbers of the particles at the interaction
vertex, similar to the usual particle-exchange Feynman diagrams. 

Both natural and unnatural parity particles can be exchanged in the
$t$-channel. The naturalness $\cal N$ for natural (${\cal N}{=}{+1}$)  and
unnatural (${\cal N}{=}{-1}$)  parity exchanges is defined as
\begin{eqnarray}
{\cal N}&=&+1 \,\,\, \mathrm{if} \,\,\, P=(-1)^J, \nonumber \\
{\cal N}&=&-1 \,\,\, \mathrm{if} \,\,\, P=(-1)^{J+1},
\end{eqnarray}
where $P$ and $J$ are the parity and spin of the exchanged particle, 
respectively.
Furthermore, in Regge theory each exchange is denoted by a signature
factor ${\cal S}{=}{\pm}1$ defined as~\cite{Irving77,Collins77}
\begin{eqnarray}
{\cal S} = P \times {\cal N} = (-1)^J,
\label{signat}
\end{eqnarray}
which enters in Eq.~(\ref{rpropa}).

\begin{table}[t]
\begin{center}
\caption{Correspondence between $t$-channel pole exchanges and the helicity
amplitudes $F_{i}$  ($i{=}1, ...,3$). Here $P$ is the parity,
$J$ the spin, $I$ the isospin, $G$ the $G$-parity, $\cal N$ the
naturalness,
and $\cal S$ the signature factor.}
\label{taba1}
\begin{tabular}{|c|c|c|c|c|c|c|c|}
\hline
$F_{i}$ & $P$  & $J$   & $I$  & $G$ & $\cal N$& $\cal S$& Exchange \\
\hline
$F_{1}$ & -1 & 1 & 1 & +1 & +1 & -1 &  $\rho$  \\
$F_{1}$ & -1 & 1 & 0 & -1 & +1 & -1 &  $\omega$  \\
$F_{2}$ & +1 &  1 & 1 & +1 & -1 & -1 & $b_1$ \\
$F_{3}$ & -1 & 1 & 1 & +1 & +1 & -1 &  $\rho$  \\
$F_{3}$ & -1 & 1 & 0 & -1 & +1 & -1 &  $\omega$  \\
\hline
\end{tabular}
\end{center}
\end{table}

In the $\gamma{p}{\to}\eta p$ reaction there is no difference between the
$\rho$ and $\omega$-exchanges if their trajectories are the same. Thus, in
some previous studies both contributions were subsummed into 
a single amplitude. However, in our study we treat $\rho$ and
$\omega$ exchanges separately, because differences in the two amplitudes 
might play role in describing
observables~\cite{Kellett70,Guidal97,Vanderhaeghen98}. 
Specifically, this allows us to account for the difference between 
$\gamma{n}{\to}\eta n$ and $\gamma{p}{\to}\eta p$ observables because for
these reactions the contributions from isoscalar and isovector exchanges
enter with different sign. The contribution of the $\rho$ and
$\omega$-exchanges to the reaction amplitudes $F_i$ are given in 
Table~\ref{taba1}, together with the relevant quantum numbers. Both $\rho$
and $\omega$ have natural parity and contribute to $F_1$ and $F_3$.

It was argued~\cite{Guidal97} that the photon beam asymmetry would be 
predominantly $\Sigma{=}$+1, if there are no other contributions besides
$\rho$/$\omega$ exchange. Note, however, that $\Sigma$ vanishes at forward 
and backward directions. This can be easily understood while considering
the
relations between the observables and the $t$-channel parity conserving
helicity amplitudes given in Refs.~\cite{Sibirtsev07,Sibirtsev08}.
Experimental data~\cite{Bussey76,Bussey81} available at photon energies
of 2.5 and 4 GeV show that the beam asymmetry depends on $t$. Therefore,
one
needs additional ingredients that contribute to the unnatural-parity
amplitudes 
$F_2$ or $F_4$.

The contribution to $F_2$ is given by the leading $b_1$ trajectory, which
we also used in the analysis of neutral and charged pion photoproduction.  
Table~\ref{taba1} shows the contribution of the $b_1$-exchange.
As we discussed in Ref.~\cite{Sibirtsev07}, the information about 
trajectories that might contribute to the $F_4$ amplitude is very poor.
Thus, in the analyses of the neutral and charged pion photoproduction we 
neglected this amplitude. For the same reason we decided to neglect 
$F_4$ also in the present study. 

The trajectories are assumed to be of linear form 
\begin{eqnarray}
\alpha(t){=}\alpha_0+\alpha^\prime{t} \ , 
\label{eq:traj}
\end{eqnarray}
where the intercept $\alpha_0$ and the slope $\alpha^\prime$ 
for the $\rho$, $\omega$  and $b_1$
trajectories are taken over from analyses of other
reactions~\cite{Sibirtsev07,Irving77,Huang10,Sibirtsev03,Sibirtsev09a}. 
In particular, for all three trajectories we adopt the slope 
$\alpha^\prime$=0.8 GeV$^{-2}$.
The values utilized for the intercepts are $\alpha_0$=0.53 for $\rho$, 
0.46 for $\omega$, and 0.51 for the $b_1$ trajectory.

The residue functions $\beta(t)$  used in our analysis are compiled in
Table~\ref{tab0}. They are similar to the ones used in some of the previous
analyses~\cite{Kellett70,Rahnama91,Sibirtsev08}. 

\begin{table}[t]
\begin{center}
\caption{Parameterization of the  residue functions $\beta(t)$ for the
amplitudes $F_i$, \,  ($i{=}1,...,3$). Here $c_ {ij}$ is the coupling
constant, where the double index refers to the amplitude $i$ and the 
type of exchange $j$, as specified in the Table.}
\label{tab0}
\renewcommand{\arraystretch}{1.1}
\begin{tabular}{|l|l|c|l|}
\hline
     & $\beta(t)$ & Exchange & $j$ \\
\hline
\multicolumn{4}{|c|}{ Pole amplitudes }\\
\hline
$F_1$ & $c_{11}$  & $\rho$ & 1 \\ 
$F_1$ & $c_{12}$  & $\omega$ & 2 \\ 
\hline
$F_2$ & $c_{23}\, t $  &
$b_1$ & 3 \\ 
\hline
$F_3$ & $c_{31} \,t $ & $\rho$ & 1 \\ 
$F_3$ & $c_{32} \,t $ & $\omega$ & 2 \\ 
\hline
\multicolumn{4}{|c|}{ Cut  amplitudes }\\
\hline
$F_1$ & $c_{14} \,   \exp[d_4t]$ & $\rho$ & 4 \\ 
$F_1$ & $c_{15} \,   \exp[d_5t]$ & $\omega$ & 5 \\ 
$F_1$ & $c_{16}\,   \exp[d_6t]$  & $b_1$ & 6 \\ 
\hline
$F_2$ & $c_{24} \, t\,  \exp[d_4t]$ & $\rho$ & 4 \\ 
$F_2$ & $c_{25} \, t\, \exp[d_5t]$ & $\omega$ & 5 \\ 
$F_2$ & $c_{26}\,  t\,  \exp[d_6t]$  & $b_1$ & 6 \\ 
\hline
$F_3$ & $c_{34}\, t\, \exp[d_4t]$ & $\rho$ & 4\\ 
$F_3$ & $c_{35} \,  t\,  \exp[d_5t]$ & $\omega$ & 5 \\ 
$F_3$ & $c_{36}\,  t\,  \exp[d_6t]$  & $b_1$ & 6 \\ 
\hline
\end{tabular}
\end{center}
\end{table}
\renewcommand{\arraystretch}{1.0}

In defining the Regge cut amplitudes we use the following parameterization
based on the absorption
model~\cite{Collins77,Kellett70,White68,White69,Henyey68}
\begin{eqnarray}
F(s,t){=} \frac{\pi \, \beta (t)}{\log{(s/s_0)}}
 \frac{1 {+}{\cal S}\exp[{-}i\pi \alpha_c (t)]}{\sin[\pi
\alpha_c(t)]\, \Gamma[\alpha_c(t)]}\!\!
\left[\frac{s}{s_0}\right]^{\alpha_c(t)-1}\!\!\!\!\!\!,
\label{eq:trajcut}
\end{eqnarray}
with the trajectories defined by
\begin{eqnarray}
\alpha_c =\alpha_0 +\frac{\alpha^\prime\alpha_P^\prime \, t}
{\alpha^\prime + \alpha_P^\prime}\,,
\label{traj2}
\end{eqnarray}
where $\alpha_0$ and $\alpha^\prime$ are taken from the pole trajectory
given by Eq.~(\ref{eq:traj}), and $\alpha_P^\prime{=}0.2$~GeV$^{-2}$ is the
slope of the pomeron trajectory. The specific form of the residue
functions 
$\beta(t)$ to be used in Eq.~(\ref{eq:trajcut}) is given in
Table~\ref{tab0}.

The relation between the differential cross section analyzed in our study
and the $t$-channel helicity amplitudes is given by 
\begin{eqnarray}
\frac{d\sigma}{dt}&=&\frac{1}{32\pi}\left[ \frac{
t|F_1|^2-|F_3|^2}{(t-4m^2)} +|F_2|^2-t|F_4|^2\right].
\label{obs1}
\end{eqnarray}

The invariant collision energy squared, $s$, and the photon 
energy $E_\gamma$ are related by 
\begin{eqnarray}
 s=m_N^2+2m_NE_\gamma, 
\end{eqnarray}
where $m_N$ is the nucleon mass.
\begin{table}[t]
\begin{center}
\caption{Parameters of the model. The $c_ {ij}$'s are the coupling
constants
for the $i$-th amplitude and the exchange of particle $j$ while the
$d_j$'s are exponents appearing in the Regge cut amplitude, 
cf. Table~\ref{tab0}.
}
\label{tabp}
\renewcommand{\arraystretch}{1.1}
\begin{tabular}{|l|c|c|c|c|}
\hline
$j$ & $c_{1j}$ & $c_{2j}$ & $c_{3j}$ & $d_j$ \\
  & $[\sqrt{\mu {\rm b}}/{\rm GeV}]$ & $[\sqrt{\mu {\rm b}}/{\rm GeV}^3]$ & 
$[\sqrt{\mu {\rm b}}/{\rm GeV}^2]$ & [GeV$^{-2}$]\\
\hline
1 & --10.71 & -- & --1.55 & -- \\
2 & --5.59 & -  & -18.05  & - \\
3 &- & 11.80  &-- & -- \\
4 & 275.47 & 30.57  & --31807 & 0.16 \\
5 & --94.71 & --35.97 &  129.14&  1.53 \\
6 & --306.89 & --32.46 & 331.94 & 0.12  \\
\hline
\end{tabular}
\end{center}
\end{table}
\renewcommand{\arraystretch}{1.0}

The parameters of our Regge model were fixed by a fit to data on 
$\eta$-meson photoproduction data published before
1982~\cite{Sibirtsev11}. 
Specifically, this concerns the differential cross sections collected at 
SLAC~\cite{Anderson70}, DESY~\cite{Braunschweig70},
Cornell~\cite{Dewire71} 
and at the 
Daresbury laboratory~\cite{Bussey76}. All these measurements were done at
and above photon energies of $E_\gamma=$2.5 GeV, which corresponds to an
invariant collision energies larger than $\sqrt{s}\simeq$2.36 GeV. 
In the actual fit only data taken at $E_\gamma{\ge}$3 GeV were used.
In addition we include into the fit procedure the
experimental results~\cite{Bussey76,Bussey81} for beam and target
asymmetries available at $E_\gamma$=3~GeV and 4~GeV. These polarization
data were collected at the Daresbury Laboratory.
The values of the model parameters are summarized in Table.~\ref{tabp}.

A comparison between those experimental
results~\cite{Anderson70,Braunschweig70,%
Dewire71,Bussey76,Bussey81} and our calculation was
presented in Ref.~\cite{Sibirtsev11}. We limited our analysis to the
range $|t|\le$ 2 GeV$^2$. Based on our experience in applying Regge
phenomenology 
to the analysis of different reactions in the past, we expect that Regge
works rather
well in this $t$ region. Furthermore, most of the data available are within
this
range.

In the near-forward direction of the $\gamma{p}{\to}\eta^0{p}$ reaction
there is an interference of the $F_1$ amplitude with the one-photon
exchange
amplitude, known as Primakoff effect~\cite{Primakoff}, which allows one 
to determine the $\eta\to\gamma\gamma$ decay width. The Regge model
described
here has been used by us in Ref.~\cite{Sibirtsev11} for a study on the
Primakoff 
effect in $\eta$-meson photoproduction. 

\section{ELSA measurements}
\begin{figure}[b]
\vspace*{-6mm}
\centerline{\hspace*{4mm}\psfig{file=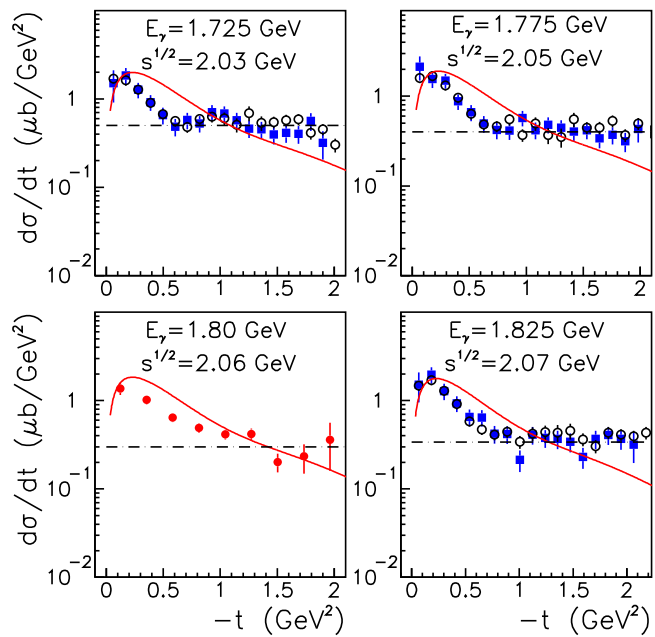,width=9.7cm}}  
\vspace*{-5mm}
\caption{\label{eta6c} 
Differential cross section for $\gamma{p}{\to}\eta p$ 
as a function of the four-momentum transfer squared at the photon energies 
$E_\gamma$ (invariant collision energies $\sqrt{s}$) indicated in the
figure. 
The data are collected at ELSA 
and published in 2005~\cite{Crede05} (filled squares),
2008~\cite{Jaegle08} (filled circles) and 2009~\cite{Crede09} (open
circles). 
The solid lines are the results of our model. The dash-dotted lines indicate
an estimation for the isotropic part of the differential cross section,
cf. text.
}
\end{figure}

\begin{figure}[t]
\vspace*{-6mm}
\centerline{\hspace*{4mm}\psfig{file=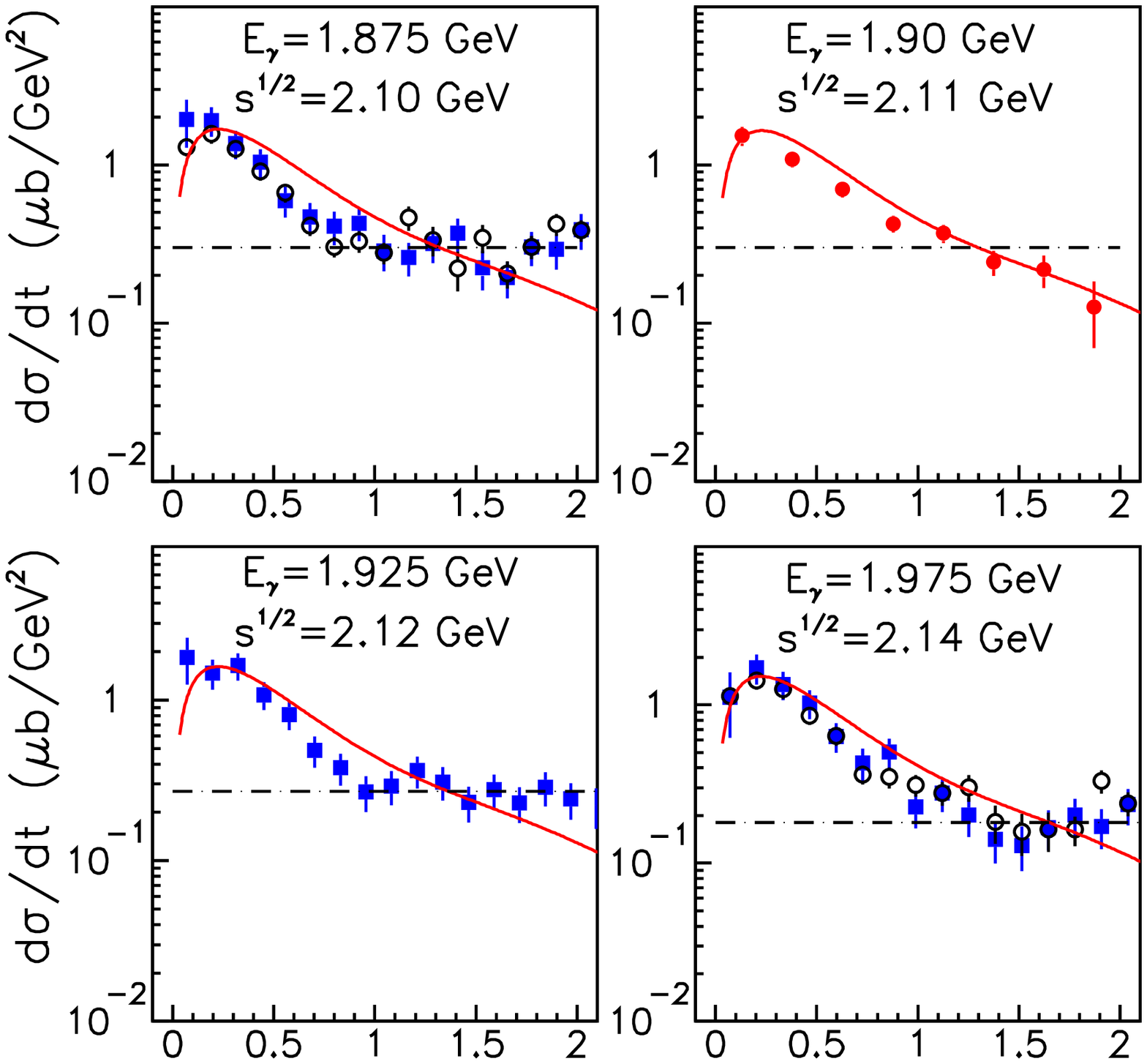,width=9.7cm}}  
\vspace*{-17mm}
\centerline{\hspace*{4mm}\psfig{file=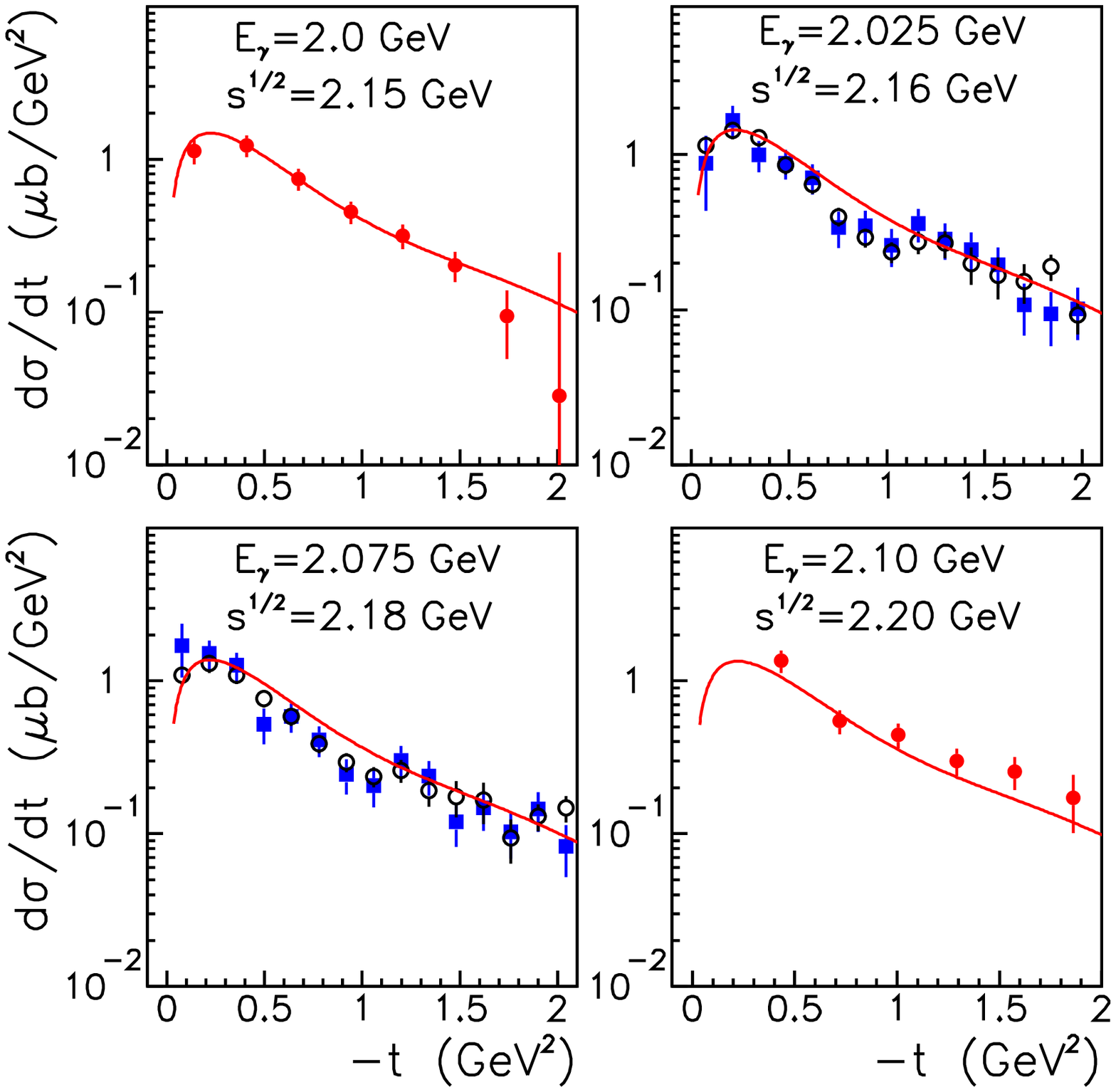,width=9.7cm}}
\vspace*{-5mm}
\caption{\label{eta6b}
Differential cross section for $\gamma{p}{\to}\eta
p$. Same notation as in Fig.~\ref{eta6c}. 
}
\end{figure}

\begin{figure}[t]
\vspace*{-6mm}
\centerline{\hspace*{4mm}\psfig{file=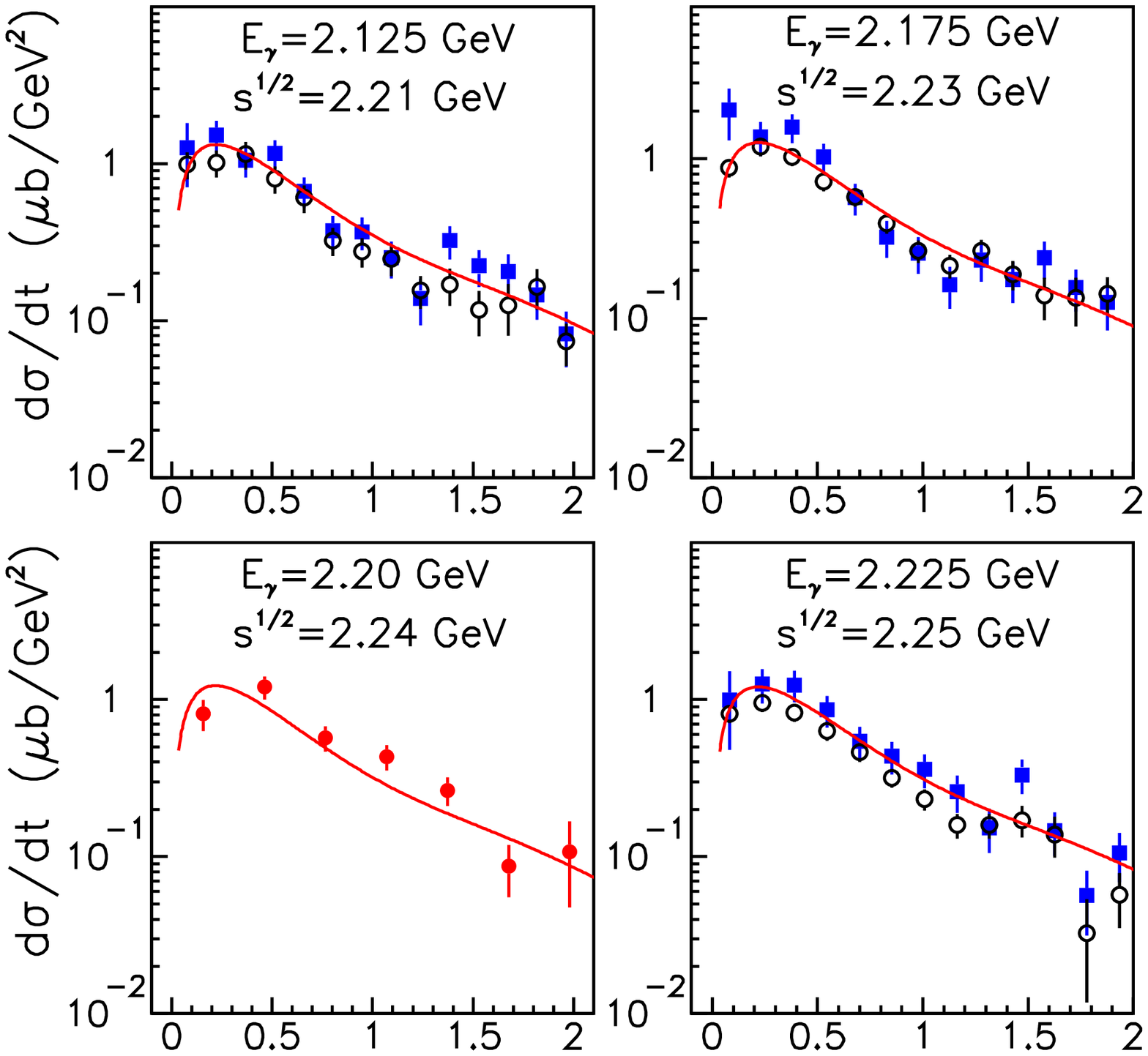,width=9.7cm}}  
\vspace*{-17mm}
\centerline{\hspace*{4mm}\psfig{file=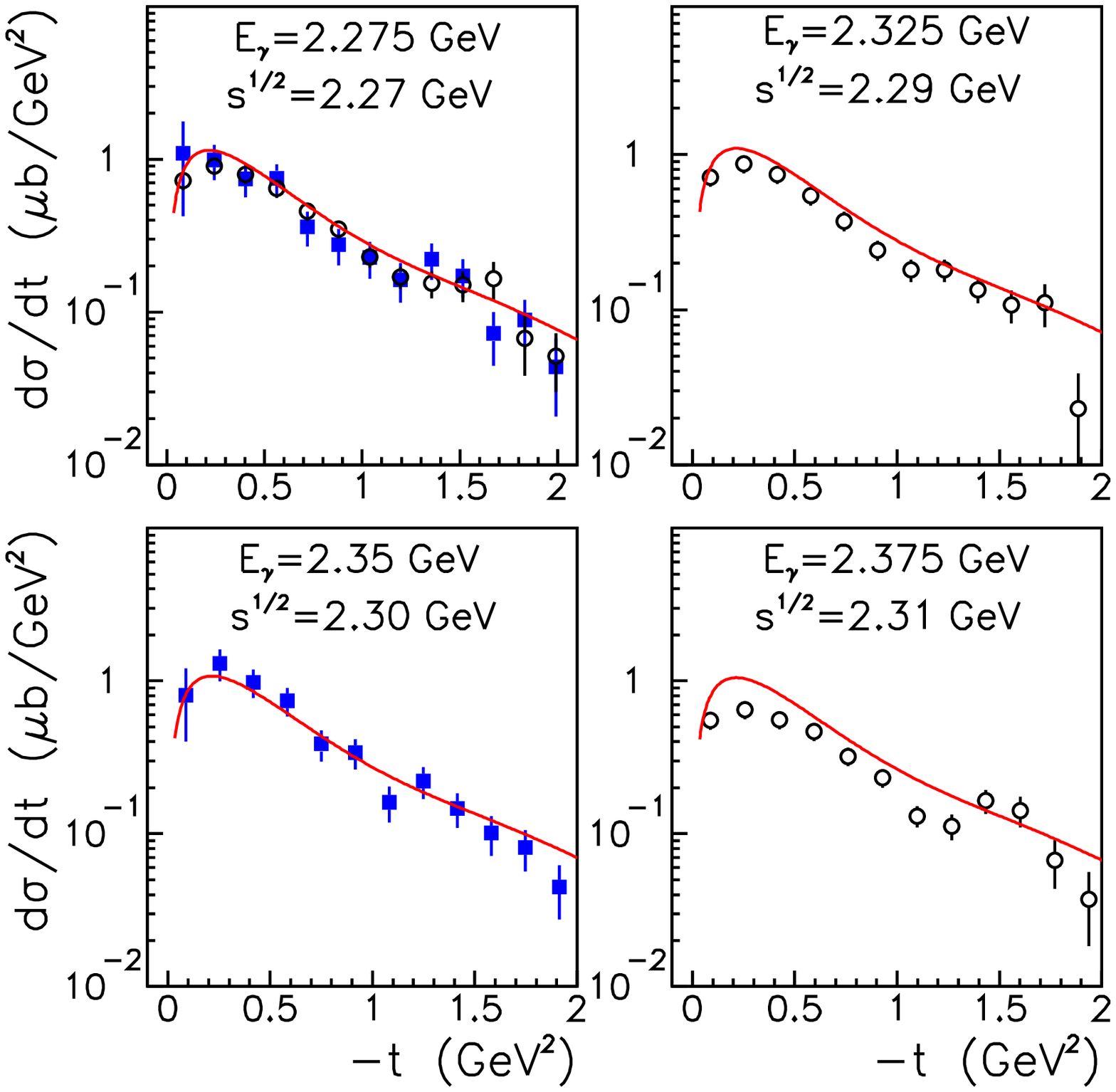,width=9.7cm}}
\vspace*{-5mm}
\caption{\label{eta6a}
Differential cross section for $\gamma{p}{\to}\eta
p$. Same notation as in Fig.~\ref{eta6c}. 
}
\end{figure}

\begin{figure}[t]
\vspace*{-6mm}
\centerline{\hspace*{4mm}\psfig{file=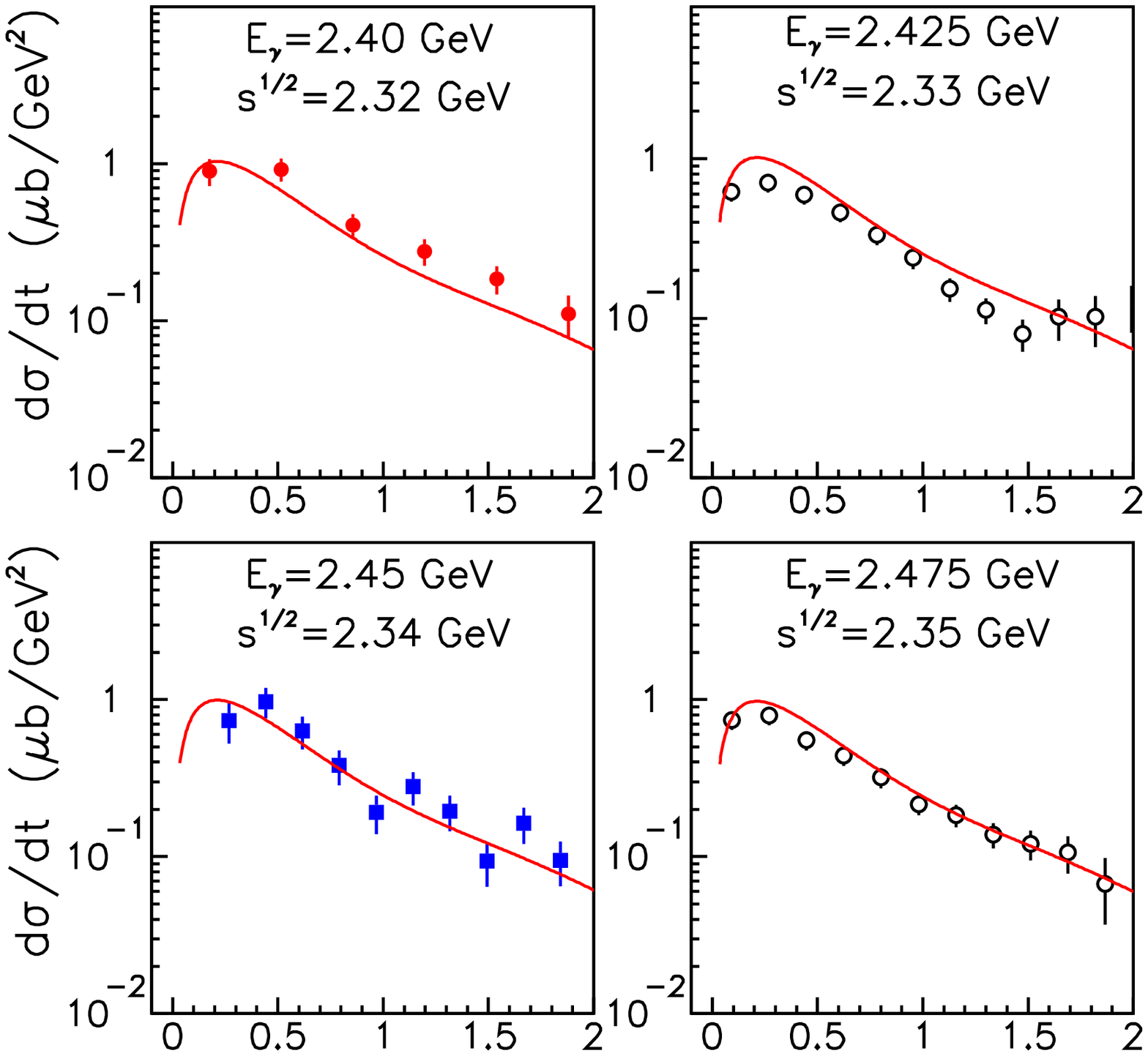,width=9.7cm}}  
\vspace*{-17mm}
\centerline{\hspace*{4mm}\psfig{file=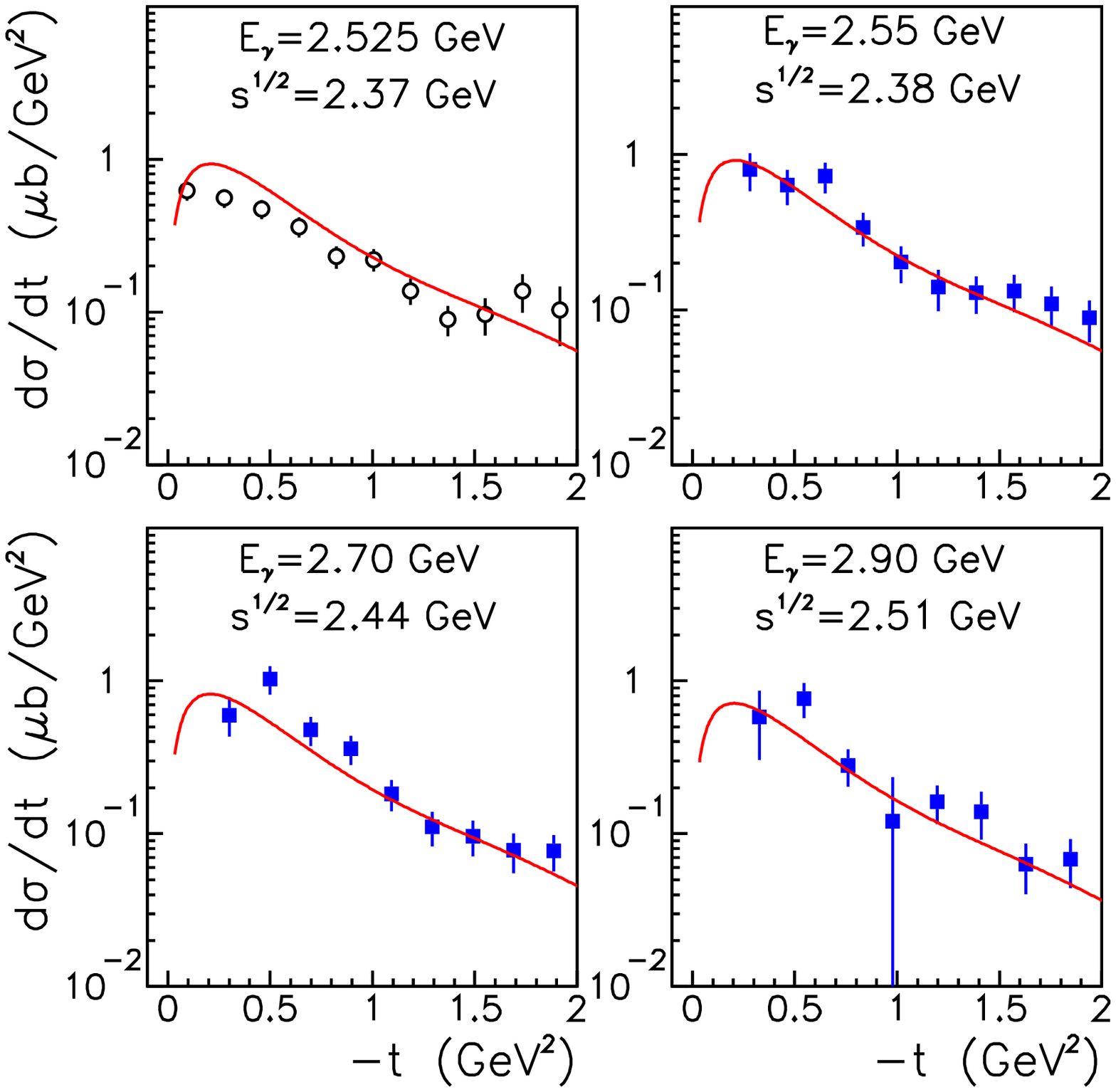,width=9.7cm}}
\vspace*{-5mm}
\caption{\label{eta6}
Differential cross section for $\gamma{p}{\to}\eta
p$. Same notation as in Fig.~\ref{eta6c}. 
}
\end{figure}

The $\gamma{p}{\to}\eta p$ differential cross section obtained at ELSA in
three different experiments~\cite{Crede09,Crede05,Jaegle08} are shown in
Figs.~\ref{eta6c}-\ref{eta6} as a function of the four-momentum
transfer squared. It can be seen that there is excellent consistency
between
the three measurements. Note that the 2005 and 2009
measurements~\cite{Crede09,Crede05} were done on a proton target,
while the data from 2008~\cite{Jaegle08} were obtained from quasi-free
photoproduction of $\eta$-mesons off a proton bound in a deuteron target. 

The results of our model are shown by solid lines. Obviously, these 
are in reasonable agreement with the data for invariant collision 
energies above 2.15 GeV, say. Note that the free parameters of 
the Regge model were fitted to $\gamma{p}{\to}\eta p$
data available at photon energies above 3 GeV. This means that the results 
shown here should be considered as predictions. Since we include only
$t$-channel
contributions, naturally one expects deviations of our results from
the experiment when going to lower and lower energies, reflecting
the increasing significance of additional contributions from 
$s$- and $u$-channel processes.

Interestingly, for energies $\sqrt{s}<$2.15 GeV the data indicate a 
peculiar feature. Namely, the
differential cross sections do not depend on the four-momentum transfer 
squared for $|t|$ above $\simeq$1.3 GeV$^2$. Such an almost isotropic
differential cross section is, in general, a sign for the 
excitation of an $s$-wave baryon resonance. The dash-dotted lines in
Figs.~\ref{eta6c} and \ref{eta6b} indicate our estimate for the 
isotropic part of the differential cross section which amounts to 
\begin{eqnarray}
 \frac{d\sigma}{dt} =0.5 \div 0.18 \,\,\, \mu\rm{b/GeV^2},
\end{eqnarray}
and decreases slowly with energy within the range 
2.03$\le\sqrt{s}\le$2.14 GeV.

The observed isotropic behaviour might be an indication for the excitation 
of the $S_{11}(2090)$ resonance. Note, however, that the latest GWU
analysis~\cite{Arndt06} of $\pi N$ scattering data found no evidence
for this resonance, despite the fact that the $\pi^-p\to\eta n$ reaction
was included in that analysis. Apparently, firm conclusions about a 
possible excitation of the $S_{11}(2090)$ baryon in the 
$\gamma p\to\eta p$ reaction require further analyses and not only
data on differential cross section but also, and more importantly,
polarization observables.

\begin{figure}[t]
\vspace*{-5mm}
\centerline{\hspace*{4mm}\psfig{file=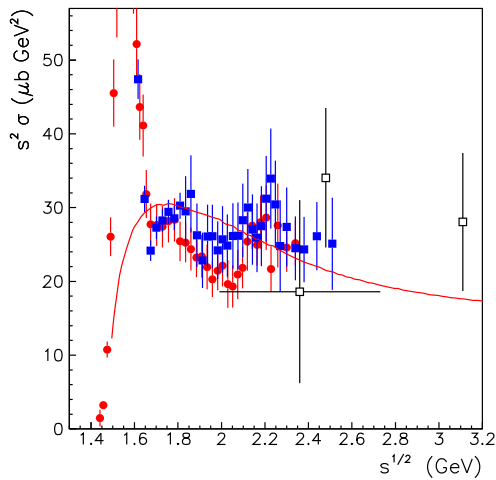,width=7.2cm}}  
\vspace*{-3mm}
\caption{\label{primet7} Total reaction cross section for $\gamma{p}{\to}\eta p$ 
as a function of the invariant collision energy. The data are from
ELSA from 2005~\cite{Crede05} (filled squares) and 2008~\cite{Jaegle08}
(filled circles). Open squares are data collected in Ref.~\cite{Landolt}.
The line is the result of our Regge model. 
The data and the model results are multiplied by $s^2$. 
}
\end{figure}
 
Finally, in Fig.~\ref{primet7} we present the total $\gamma{p}{\to}\eta p$ 
reaction cross section as a function of the invariant collision energy
$\sqrt{s}$.
To avoid a presentation with logarithmic scale we have multiplied the data
as well as the model results with the leading Regge exponent ($s^2$). 
In case of the ELSA experiment the reaction cross section was
obtained by summing over the angular bins and extrapolating outside the
angular range of the actual measurement~\cite{Crede05}. 
The filled squares show the ELSA results from 2005~\cite{Crede05} while 
the filled circles are those from 2008~\cite{Jaegle08}. 
Open squares are pre-2000 data collected in Ref.~\cite{Landolt}.

In view of the very large uncertainties of the ELSA results
from 2005~\cite{Crede05} it is difficult to draw conclusions 
from the comparison between those data and the model calculation. 
However, it is obvious that the model results are at variance with the 
ELSA data from 2008~\cite{Jaegle08} for invariant collision energies 
around and below 2.15 GeV. This observation is in line with the 
shortcomings of the model with regard to the differential cross sections,
discussed above. 

In summary, we found mutual consistency between the 
$\gamma{p}{\to}\eta p$ measurements~\cite{Crede09,Crede05,Jaegle08}
available from ELSA. We observe a reasonable agreement between
the results obtained in these three different ELSA experiments and our
predictions at invariant collision energies above 2.15 GeV or photon
energies $E_\gamma{\ge}2$ GeV. Since the free parameters of our model were
fixed in a fit to the pre-2000 data available at photon energies
above 3 GeV,
we conclude that there is also consistency between the ELSA data and 
the previous measurements performed at higher 
energies~\cite{Anderson70,Braunschweig70,Dewire71,Bussey76,Bussey81}.

\section{CLAS measurements}
The CLAS Collaboration has performed two different experiments for
$\gamma{p}{\to}\eta p$~\cite{Williams09,Dugger02} at JLab. 
The data published in 2002~\cite{Dugger02} cover invariant
energies up to 2.12 GeV, while the most recent
measurement~\cite{Williams09} 
extends up to $\sqrt{s}\simeq$2.79 GeV.

\begin{figure}[t]
\vspace*{-6mm}
\centerline{\hspace*{4mm}\psfig{file=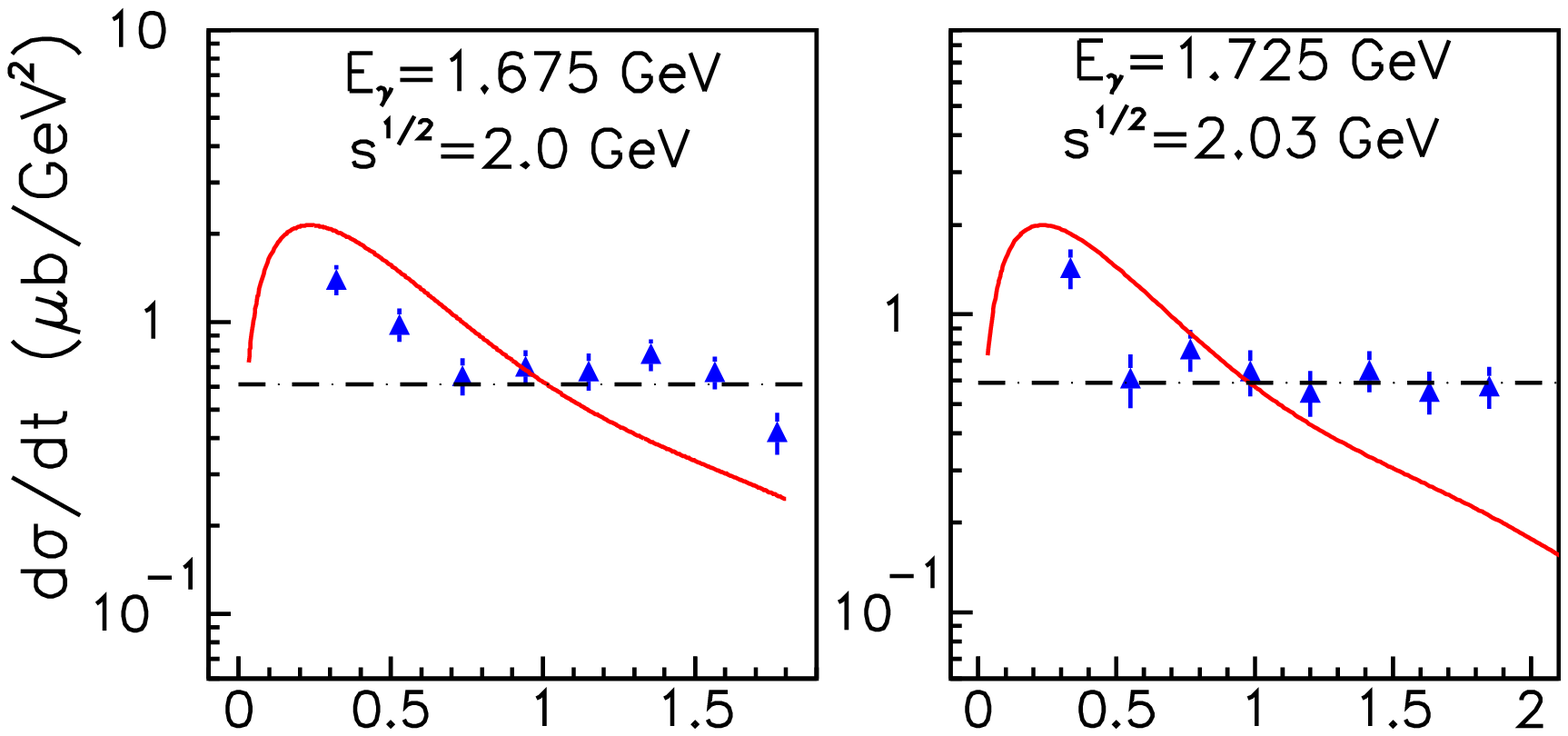,width=9.7cm}}  
\vspace*{-58mm}
\centerline{\hspace*{4mm}\psfig{file=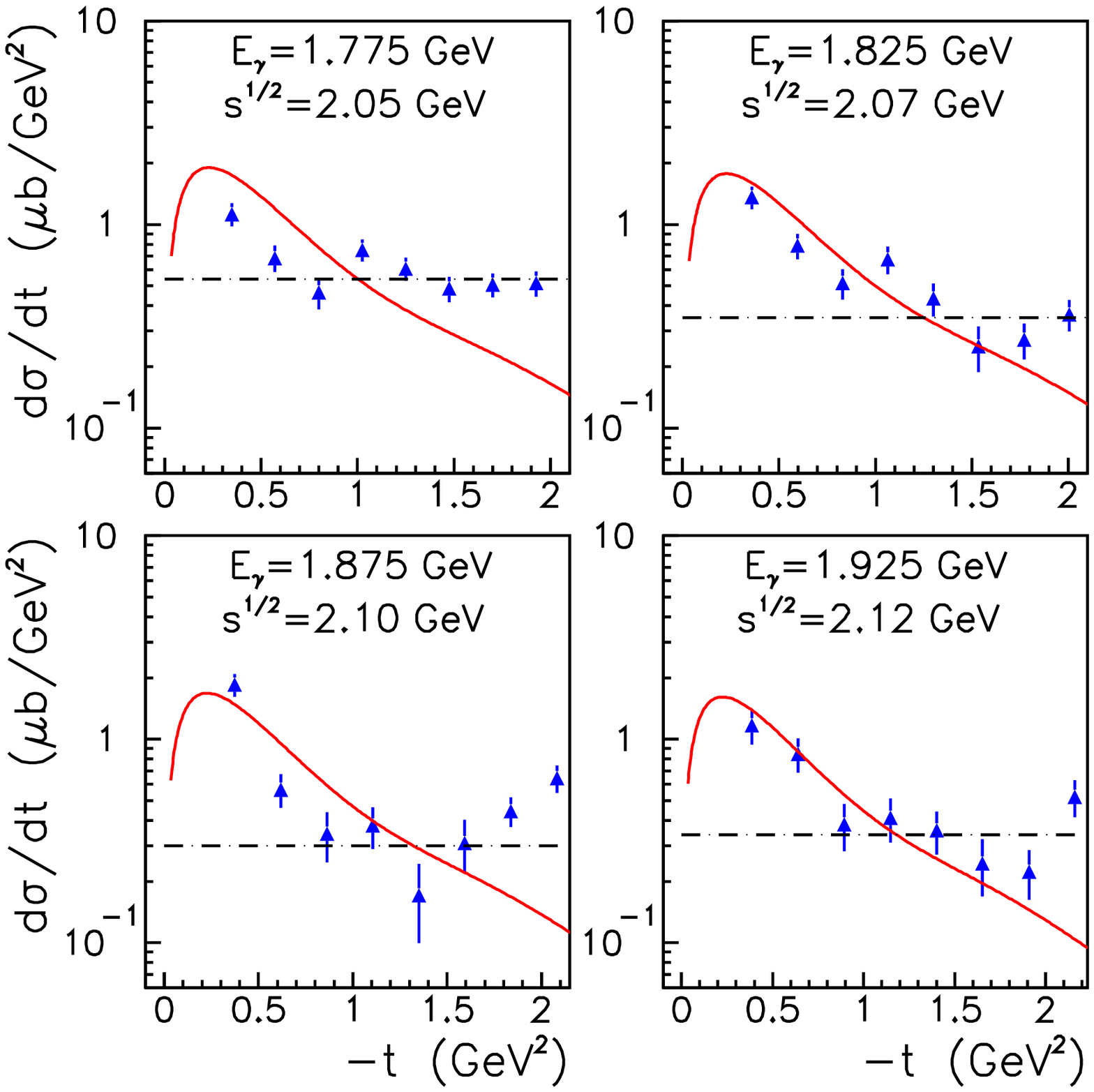,width=9.7cm}}  
\vspace*{-5mm}
\caption{\label{ceta1}
Differential cross section for $\gamma{p}{\to}\eta p$ 
as a function of the four-momentum transfer squared at the photon energies 
$E_\gamma$ (invariant collision energies $\sqrt{s}$) indicated in the
figure. 
Filled triangles represent data by the CLAS Collaboration published in 2002~\cite{Dugger02}. 
The solid lines are the results of our model. 
The dash-dotted lines indicate an estimation for the isotropic part of the 
differential cross section, cf. text.
}
\end{figure}

\begin{figure}[t]
\vspace*{-6mm}
\centerline{\hspace*{4mm}\psfig{file=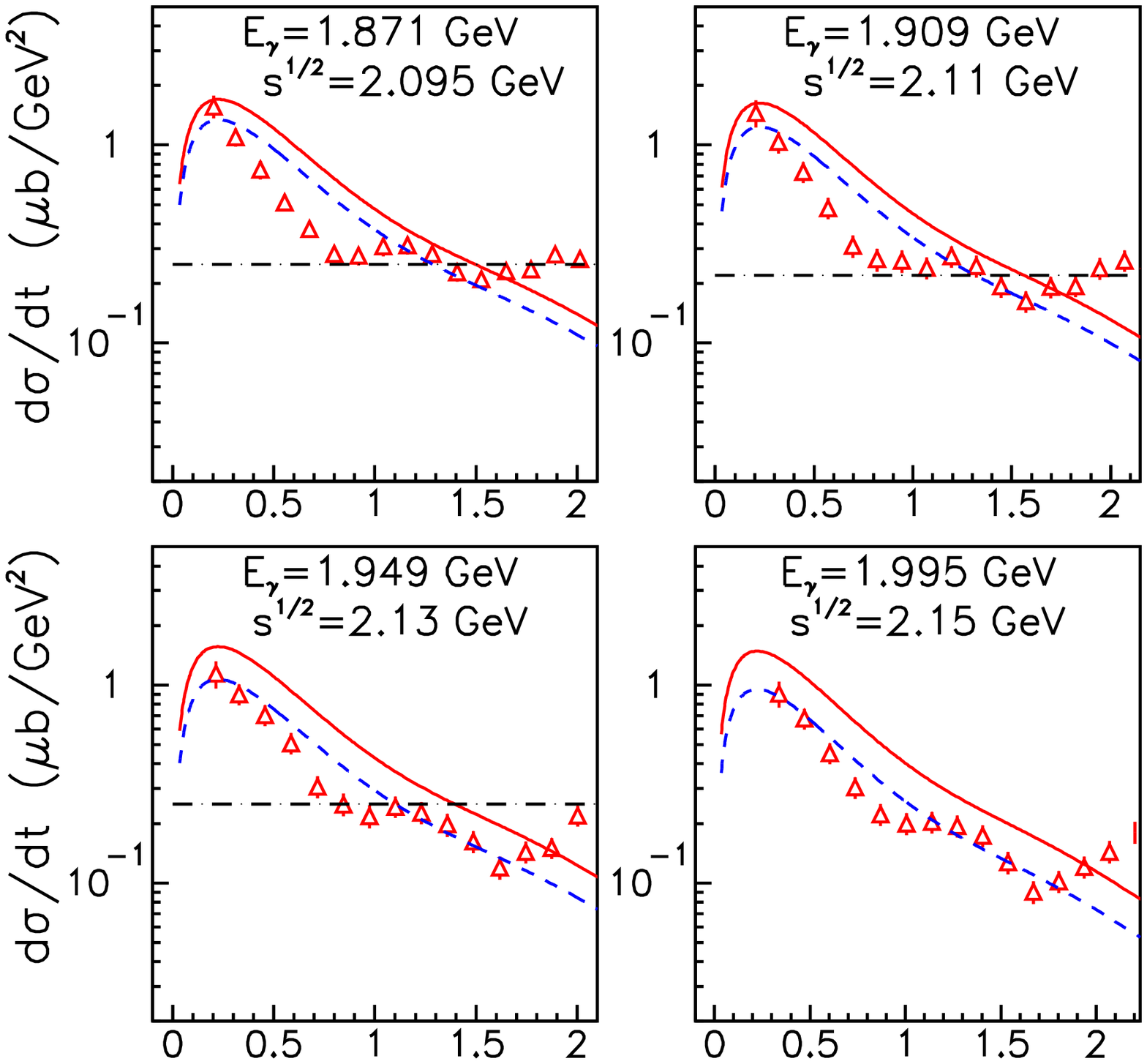,width=9.7cm}}  
\vspace*{-18mm}
\centerline{\hspace*{4mm}\psfig{file=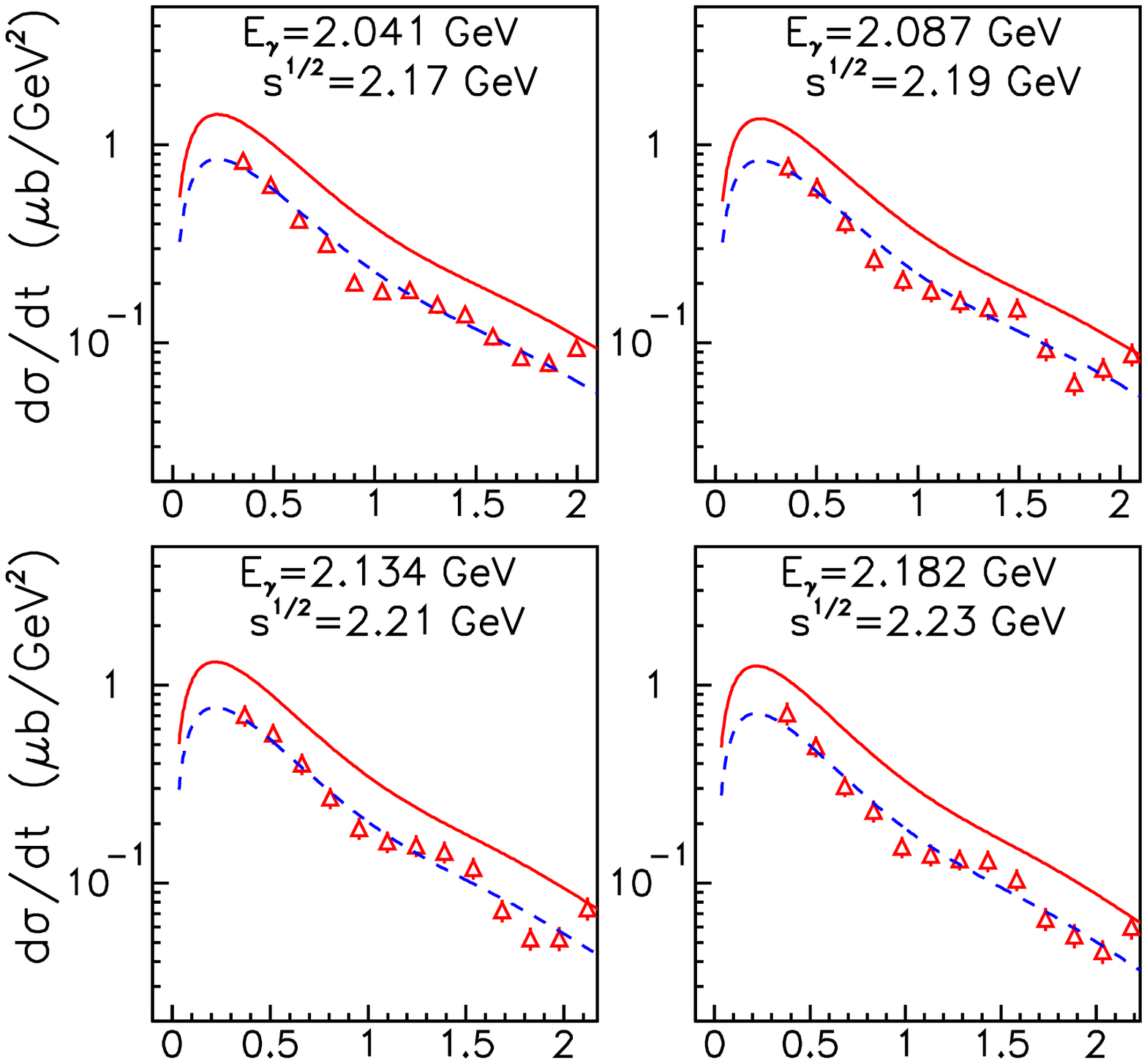,width=9.7cm}}  
\vspace*{-5mm}
\caption{\label{ceta4}
Differential cross section for $\gamma{p}{\to}\eta p$ 
as a function of the four-momentum transfer squared at the photon energies 
$E_\gamma$ (invariant collision energies $\sqrt{s}$) indicated in the
figure. 
Open triangles represent data by the CLAS Collaboration published in 2009~\cite{Williams09}. 
The solid lines are the results of our model. When allowing the normalization to float
the dashed lines are obtained. 
The dash-dotted lines shown in the plots for $E_\gamma \le$ 1.949 GeV indicate an 
estimation for the isotropic part of the differential cross section, cf. text. 
}
\end{figure}

In Fig.~\ref{ceta1} we present the data from the 2002
measurement~\cite{Dugger02}
together with our model results. In view of the disagreement with the
ELSA data below $\sqrt{s}\approx$2.15 GeV, reported in the last section, 
it is not surprising that the model prediction agrees only at the highest
energy ($\sqrt{s}$=2.12 GeV) and for $|t|{\le}$2 GeV$^2$ roughly with the CLAS results,
considering the large experimental uncertainties. At lower energy the
$\gamma{p}{\to}\eta p$ differential cross sections show practically no
$t$-dependence\footnote{Note that at large $|t|$ or small $|u|$
the experimental differential cross sections indicate some increase with
$u\to0$, which is typical for contributions from $u$-channel processes.} 
for $|t|$ above $\simeq$1.3 GeV$^2$. We illustrate this observation by the
dash-dotted lines in the Fig.~\ref{ceta1}. These finding are in line with those
for the ELSA data, discussed in the previous section. 

\begin{figure}[t]
\vspace*{-6mm}
\centerline{\hspace*{4mm}\psfig{file=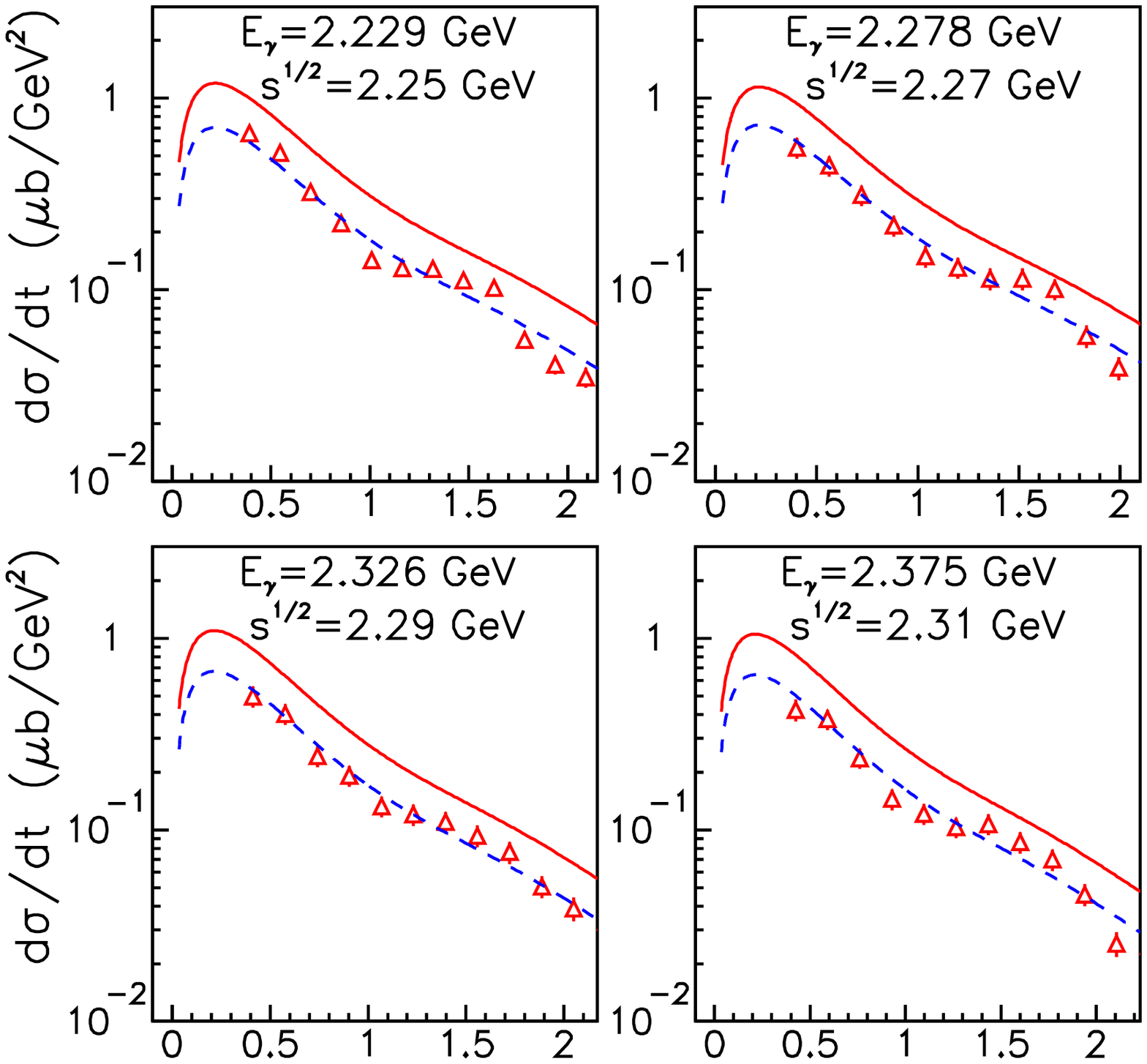,width=9.7cm}}  
\vspace*{-18mm}
\centerline{\hspace*{4mm}\psfig{file=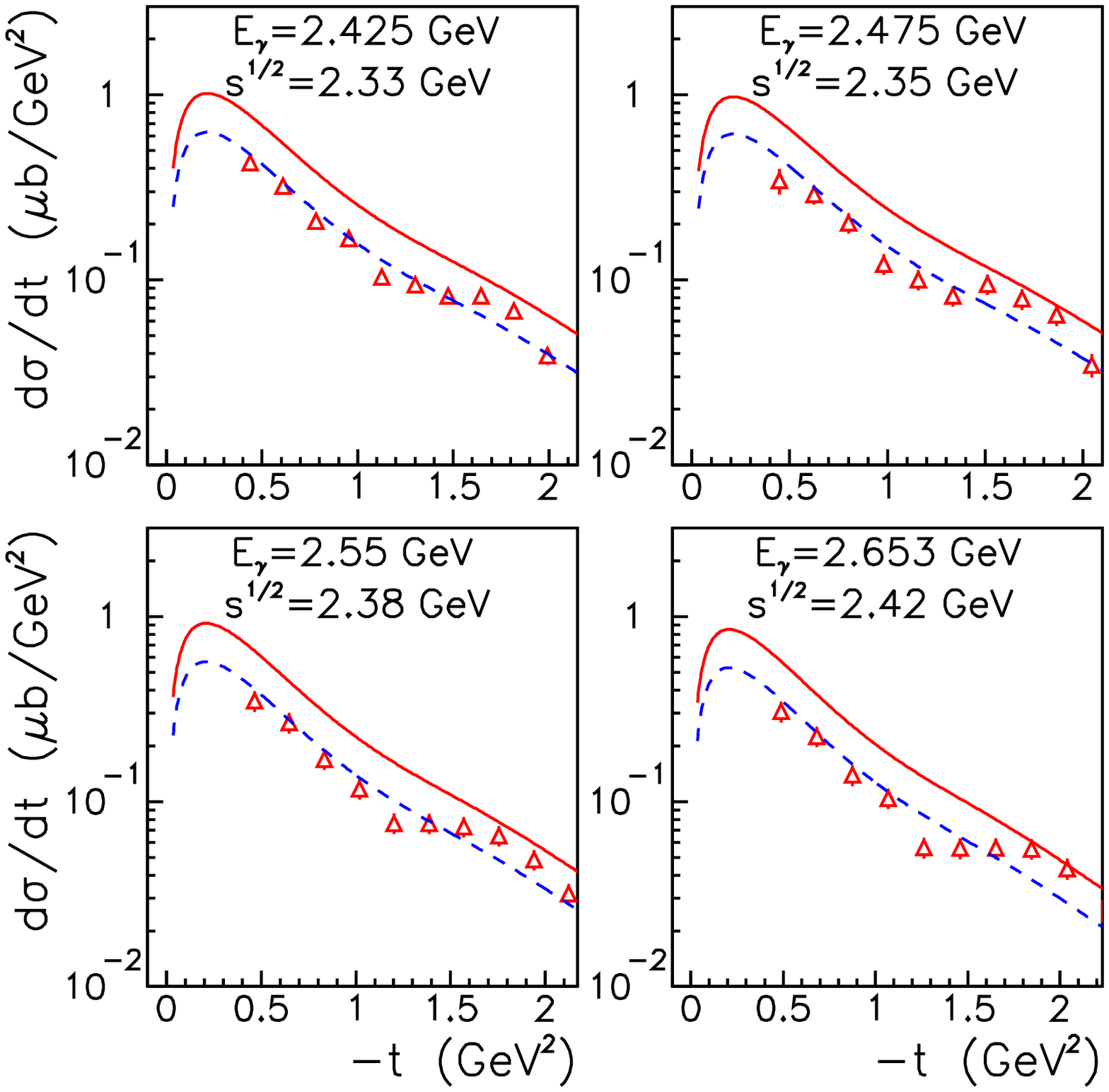,width=9.7cm}}  
\vspace*{-5mm}
\caption{\label{ceta3}
Differential cross section for $\gamma{p}{\to}\eta
p$. Same notation as in Fig.~\ref{ceta4}. 
}
\end{figure}

\begin{figure}[t]
\vspace*{-6mm}
\centerline{\hspace*{4mm}\psfig{file=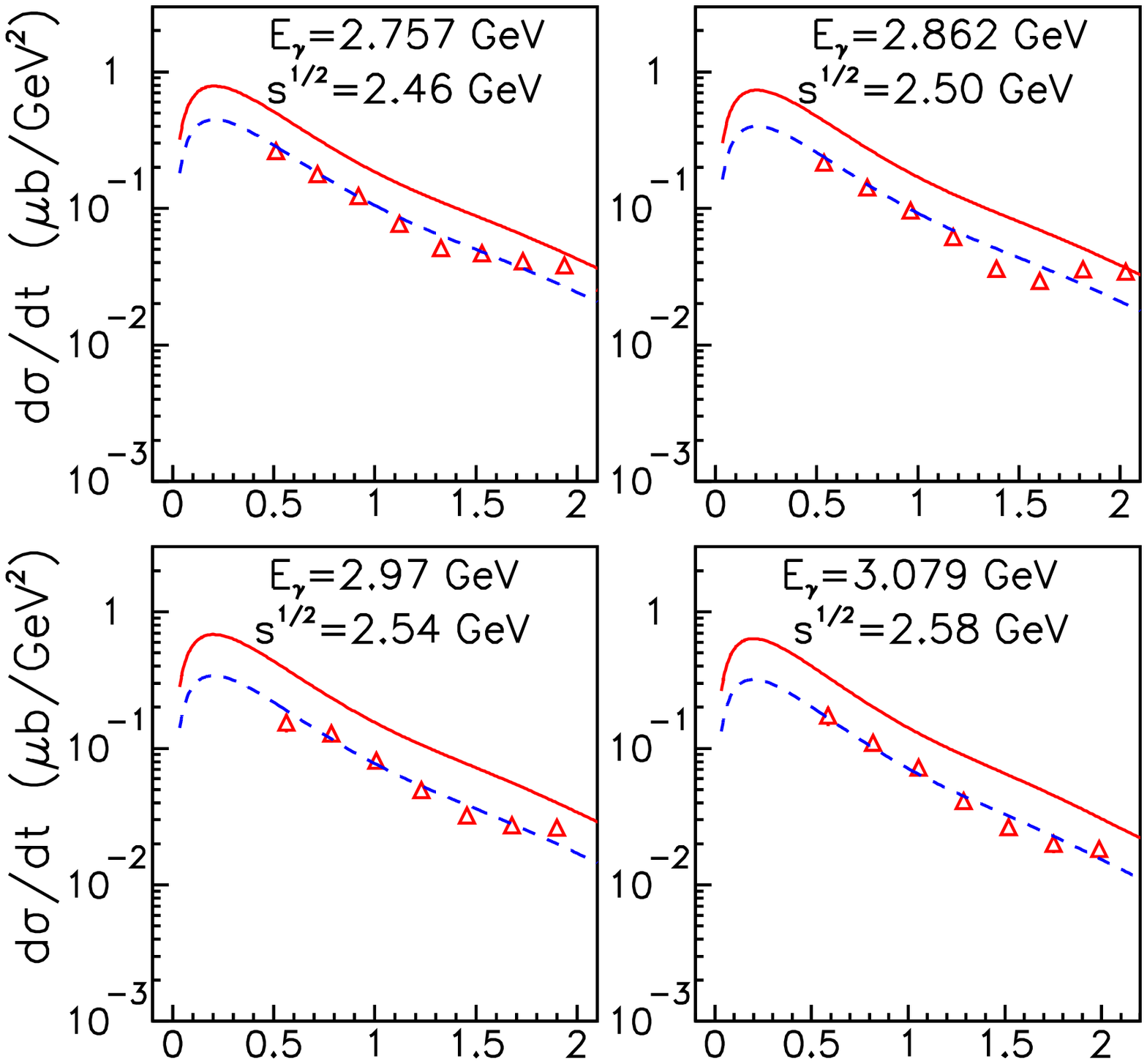,width=9.7cm}}  
\vspace*{-18mm}
\centerline{\hspace*{4mm}\psfig{file=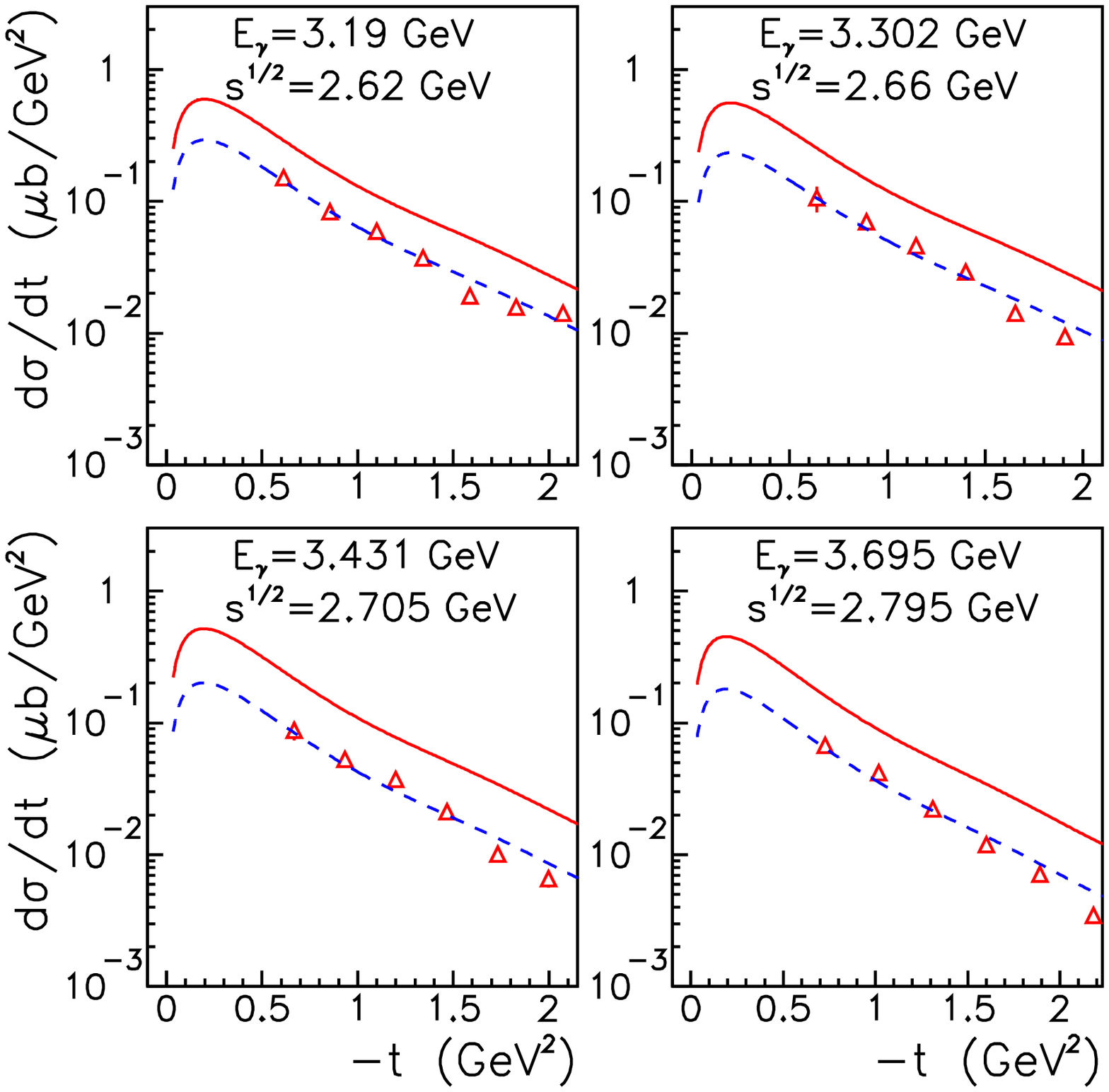,width=9.7cm}}  
\vspace*{-5mm}
\caption{\label{ceta2}
Differential cross section for $\gamma{p}{\to}\eta
p$. Same notation as in Fig.~\ref{ceta4}}. 
\end{figure}

The most recent results from CLAS~\cite{Williams09} are displayed in
Figs.~\ref{ceta4}-\ref{ceta2}. The plotted values for the data (triangles) 
are taken from the Durham data base~\cite{Durham}. An additional
error of 11\% in average was included in quadrature following the results
given in Table 1 of Ref.~\cite{Williams09}. It is clear that the
experimental results are in strong disagreement with our calculation which
are indicated by the solid lines. The discrepancy is especially surprising 
for invariant energies above 2.54 GeV or $E_\gamma>$ 3 GeV. As said before,
the 
parameters of our model were fixed~\cite{Sibirtsev11} utilizing the data on
differential cross sections and polarization available at photon energies
above 3 GeV. Thus, the observed disagreement means that the new CLAS
results are inconsistent with the pre-2000
measurements~\cite{Anderson70,Braunschweig70,Dewire71,Bussey76,Bussey81}
from SLAC, DESY, Cornell and the Daresbury laboratory. But they are also
inconsistent with the three sets of 
measurements~\cite{Crede09,Crede05,Jaegle08} from ELSA, and the 2002 
results from the CLAS collaboration itself~\cite{Dugger02}.

\begin{figure}[t]
\vspace*{-6mm}
\centerline{\hspace*{4mm}\psfig{file=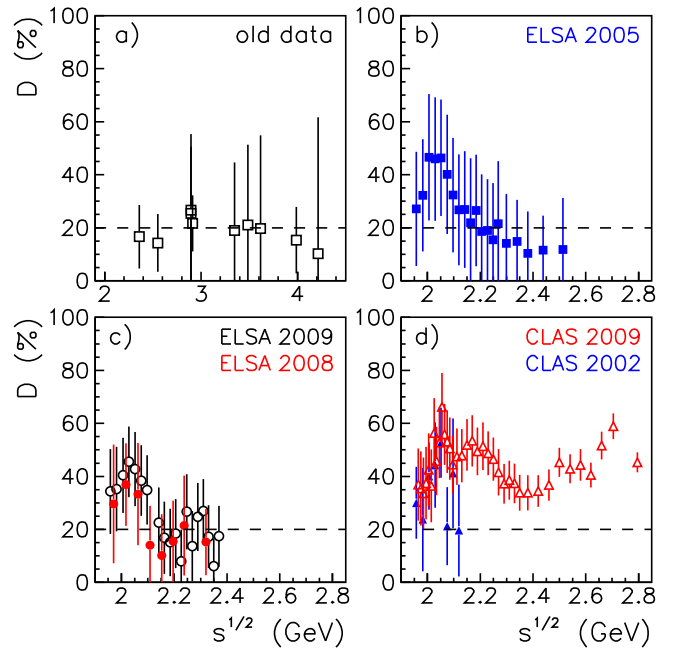,width=9.7cm}}
\vspace*{-4mm}
\caption{\label{eta2a}The deviation $D$ as a function of invariant 
collision energy $\sqrt{s}$ shown for different experiments: pre-2000
measurements (open squares), ELSA measurements from 2005~\cite{Crede05}
(filled squares), 2008~\cite{Jaegle08} (filled circles), 
2009~\cite{Crede09} (open circles) and  CLAS measurements from
2002~\cite{Dugger02} (filled triangles), 2009~\cite{Williams09} (open
triangles). The dashed lines indicate an averaged $D$ value obtained 
with pre-2000 data. }
\end{figure}

\section{Comparison of the different data sets}

A direct comparison of the different measurements is difficult because, in
general,
the ELSA and CLAS data are available at different energies and different
angles.
In the present study we circumvent this problem by comparing the
experimentally
observed differential cross sections with the predictions of our Regge 
model~\cite{Sibirtsev11}. 

For the following discussion let us define a function $D$
as a measure of the discrepancy (or deviation) of the experimental results 
for the $\eta$-meson photoproduction differential 
cross section from our model results:
\begin{eqnarray}
D (\sqrt{s}) = \frac{1}{N}
\sum_{i=1}^N 
\frac{| \,(d\sigma/dt)^{\exp}_i-(d\sigma/dt)^{\rm th}_i\, |} 
{(d\sigma/dt)^{\exp}_i} \ . 
\label{dev}
\end{eqnarray}
The summation is done over all data points in the range of the
four-momentum 
transfer squared $t{\ge}$-2 GeV$^2$ at fixed invariant collision energy
$\sqrt{s}$,
and $N$ is the number of experimental points at each energy.
Here $(d\sigma/dt)^{\exp}_i$ stands for the experimental value  
while $(d\sigma/dt)^{\rm th}_i$ denotes the result of our Regge model 
for the $i$th data point at a specific energy $\sqrt{s}$. In practice,
Eq.~(\ref{dev}) represents the relative uncertainty of our model with
respect to the data averaged over the $t$-distribution or, alternatively,
over the angular spectrum at a fixed energy. We do not include polarization
data since there are only few data points and, moreover, 
the data at two photon energies are afflicted by large uncertainties.

\begin{figure}[b]
\vspace*{-6mm}
\centerline{\hspace*{0mm}\psfig{file=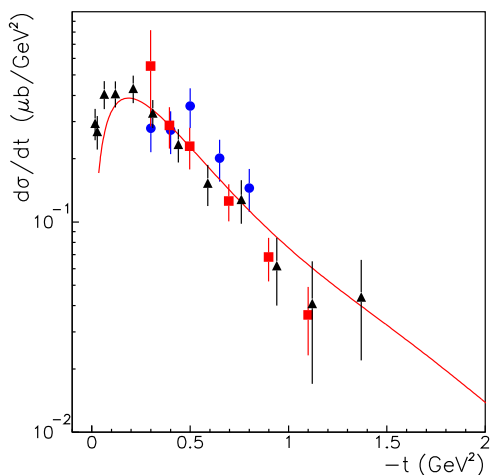,width=7.2cm}}
\vspace*{-4mm}
\caption{\label{eta5}
Differential cross section for $\gamma{p}{\to}\eta p$ 
as a function of the four-momentum transfer squared at the photon energy
of 4 GeV.
The squares represent data from SLAC~\cite{Anderson70}, 
triangles are from DESY~\cite{Braunschweig70} and circles are from
Cornel~\cite{Dewire71}. 
The solid line is the result of our model.
}
\end{figure}

To indicate the uncertainties of the experimental values we consider a 
relative error averaged over the $t$-spectrum at each energy, which is 
given by
\begin{eqnarray}
 \delta D (\sqrt{s}) =  \frac{1}{N}
\sum_{i=1}^N 
\frac{\sqrt{(\Delta^{\rm stat}_i)^2+(\Delta^{\rm sys}_i)^2}} 
{(d\sigma/dt)^{\exp}_i} \ . 
\label{edev}
\end{eqnarray}
Here $\Delta^{\rm stat}$ and $\Delta^{\rm sys}$ are the absolute statistical and
systematical experimental uncertainties given for the $i$th data point. 
They are combined in quadrature similar to the pion data analysis performed
in Ref.~\cite{Arndt06}. 

It is useful to consider another quantity that was
introduced in Refs.~\cite{Adelseck90,Mart07} for the evaluation of the
consistency of
data for the $\gamma{p}\to K^+\Lambda$ reaction. The deviation of 
each data point from the model result was computed from 
\begin{eqnarray}
R_i  = \frac{(d\sigma/dt)^{\exp}_i -
(d\sigma/dt)^{\rm th}_i}{\Delta\sigma^{\rm stat}_i},
\end{eqnarray}
with the mean value $\langle R \rangle$
and second algebraic moment $\langle R^2 \rangle$ defined by
\begin{eqnarray}
 \langle R \rangle = \frac{1}{N} \sum_{i=1}^N R_i, \\
 \langle R^2 \rangle = \frac{1}{N} \sum_{i=1}^N R_i^2 =\frac{\chi^2}{N}.
\label{chi2}
\end{eqnarray}
As indicated, the quantity $\langle R^2 \rangle$ is equivalent to the
standard
$\chi^2$ per data point. 
The evaluation of $R_i$ involves the statistical uncertainty of the data
points, which can be very different for different measurements. Thus, 
it can happen that the relative deviation is large, but due to small  
values of $\Delta\sigma^{\rm stat}_i$ and not because of a
substantial difference between model predictions and the data.
This has to be taken into account in the interpretation of the results for
$R_i$, etc. In the present study we will show the energy dependence of the
second 
moment $\langle R^2 \rangle$ in order to illustrate the role of the
statistical
uncertainties of the experimental results.

Results for $D(\sqrt{s})$, evaluated for the pre-2000
data~\cite{Anderson70,Braunschweig70,Dewire71,Bussey76,Bussey81} on
$\gamma{p}{\to}\eta p$, are shown in Fig.~\ref{eta2a}a). 
Note that the deviation presented in Fig.~\ref{eta2a} is given in percent. 
From the results in Fig.~\ref{eta2a}a) we conclude that the
deviation of the model calculation from those pre-2000 experimental
results 
amounts to about 20\% in average. The dashed line represents this average value 
and is shown in all panels in Fig.~\ref{eta2a} for illustrative purposes. 
This deviation is well within the uncertainties of the experimental data 
as is reflected by the fact that the error bars, corresponding to the variation
$\delta D$, cross the zero-percent line in most of the cases. 
 
One could believe that the deviation $D$ can be further minimized, e.g., by 
renormalizing the model 
results\footnote{Actually we use a $\chi^2$ minimization procedure to find an 
optimal description of the data. In this case we also account for the
uncertainty of each experimental point. Note that the definition of the 
$\chi^2$ differs from that of the quantity $D$.}. 
But this is not the case because from a certain level onwards $D$ provides 
only a measure for the reproduction of the $t$-dependence. 
This is well illustrated by Fig.~\ref{eta5} where the differential cross section
for $\gamma{p}{\to}\eta p$ at the photon energy of 4 GeV is displayed. 
Indeed an overall normalization would not improve the description of
the data since the $t$-dependence is an essential feature of the model
as well as of the data. 

\begin{figure}[t]
\vspace*{-6mm}
\centerline{\hspace*{4mm}\psfig{file=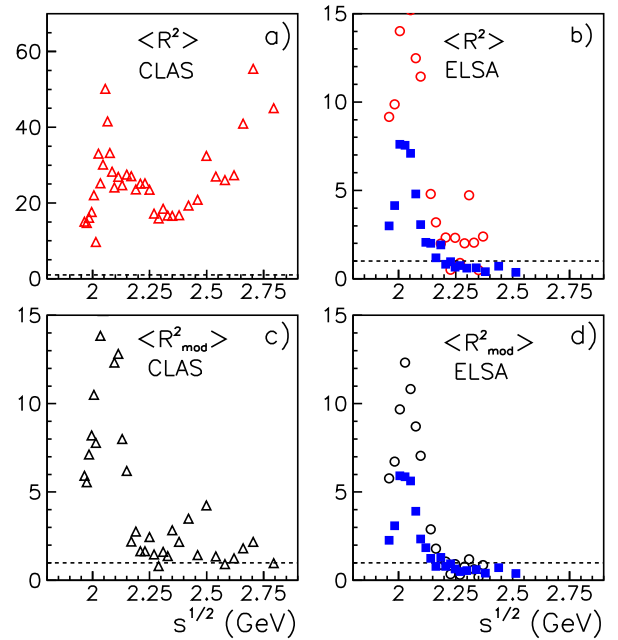,width=9.7cm}}  
\vspace*{-5mm}
\caption{\label{ceta5}
The second moment of the deviation $\langle R^2 \rangle$,
cf. Eq.~(\ref{chi2}), evaluated for the 2009 CLAS 
data~\cite{Williams09} (open triangles) and the ELSA data from
2009~\cite{Crede09} (open circles) and 2005~\cite{Crede05} (filled squares). 
The dashed lines indicate $\langle R^2 \rangle{=}1$. 
In panels a) and b) we show results for the data as published, 
while in c) and d) the modified $\langle R^2_{\rm mod} \rangle$ that involves
the renormalization factor $X$ are presented.
}
\end{figure}

Fig.~\ref{eta2a}b) summarises the deviation between our calculation and
the results from the ELSA experiment published~\cite{Crede05} in 2005.  
The values of the cross sections and the error bars were taken from the 
Durham data base. In the evaluation
of the relative uncertainties $\delta D$ we include the statistical and
systematic errors as given in the Tables in Refs.~\cite{Crede05,Durham} 
and, in addition, we add in quadrature a normalization error of 15\%.

The agreement between the data and the model at energies
$\sqrt{s}>$2.15~GeV is quite reasonable, as expected from the comparison
presented in Sect. 3. At these energies the relative
deviation amounts in average to $\simeq$7\%, which is compatible with the
experimental uncertainties.  For lower energies the deviation increases
and it reaches a maximum value of about 45\% around invariant energies of 2.1
GeV. Interestingly, the result for $D$ resembles pretty much a distribution
one would expect for a resonance structure.

Results for
the more recent data from ELSA~\cite{Jaegle08,Crede09} published in 2008
and 2009 are summarized in Fig.~\ref{eta2a}c). Again the values for the 
differential cross section and the statistical and systematic
uncertainties were taken from data base~\cite{Durham}. These errors were
added quadratically to obtain the total error as indicated by
Eq.~(\ref{edev}). The systematic uncertainty of the 2008 
data~\cite{Jaegle08} was assumed to be 10\% at photon energies below 2 GeV 
and 15\% at higher photon energies.
We find that there is also a good overall agreement between the model calculation 
and both sets
of the recent ELSA measurements at 2.15$<\sqrt{s}<$2.4 GeV. At these energies
the relative deviation amount to an average of $\simeq$5\%, which is less
than the experimental uncertainties. For lower energies the deviation increases
and reaches values around 40\% for $\sqrt{s}\simeq$2.1 -- similar to what was found
for the 2005 data from the ELSA collaboration.  

\begin{figure}[t]
\vspace*{-6mm}
\centerline{\hspace*{4mm}\psfig{file=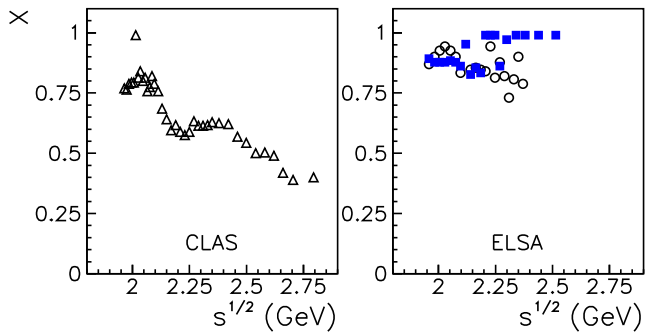,width=9.7cm}}  
\vspace*{-45mm}
\caption{\label{ceta5a} Renormalization factor $X$ as a function
of the invariant collision energy. Open triangles represent the values 
obtained in the analysis of the CLAS data~\cite{Williams09}, while the open 
circles and filled squares are for the ELSA measurements from 2009~\cite{Crede09}
and 2005~\cite{Crede05}, respectively.
}
\end{figure}

Finally, in Fig.~\ref{eta2a}d) the deviation $D$ between our model result and the 
experimental data~\cite{Dugger02,Williams09} from CLAS is shown.
With respect to the 2002 data~\cite{Dugger02} the
overall normalization uncertainty was estimated to be in the range of 
3\% to 7\%, increasing with photon energy. For the data published in
2009~\cite{Williams09} an additional average error of 11\% was included
in quadrature following the results given in Table~1 of Ref.~\cite{Williams09}.

The few data points from the 2002 measurements~\cite{Dugger02} 
at invariant collision energies above $\simeq$2.1 GeV 
are reasonably well described by the model. However, the very 
recent CLAS results~\cite{Williams09} deviate substantially from the model 
over the whole range of 2.1${<}\sqrt{s}{<}2.8$ GeV, {\it i. e.} for energies 
where we observe good agreement with all other experimental results. 
Taking into account the systematic uncertainties shows that the two CLAS 
measurements~\cite{Dugger02,Williams09} themselves are not consistent with 
each other. 

In the following let us investigate whether the observed significant discrepancy 
with the new CLAS data could be resolved by changing the absolute normalization 
of those data. In fact, the results displayed in Figs.~\ref{ceta4}-\ref{ceta2} 
suggest that for $\sqrt{s}\ge$2.54 GeV the shape of the $t$-dependence exhibited 
by the CLAS data looks indeed very similar to our model predictions. 
For that purpuse we consider a modified $\chi^2$ function given by
\begin{eqnarray}
\chi^2_{mod}   = \sum_i^N \left[ \frac{(d\sigma/dt)^{\exp}_i -
X(d\sigma/dt)^{\rm th}_i}{\Delta\sigma^{\rm stat}_i}\right]^2
\label{chi2m}
\end{eqnarray}
and evaluate the renormalization factor $X$ at each energy by searching for the
minumum of $\chi^2_{\rm mod}$. The quantity
$\chi^2_{\rm mod}$ should be compared with the second moment of the deviation 
$\langle R^2 \rangle$
given by Eq.~(\ref{chi2}), specifically, we can introduce
$\langle R^2_{\rm mod}\rangle = \chi^2_{\rm mod} / N$. 

The open triangles in Fig.~\ref{ceta5}a) show $\langle R^2 \rangle$ as a function
of the invariant collision energy evaluated for the 2009
data~\cite{Williams09} from CLAS. Apparently the discrepancy between our
model and the data is huge, even at $\sqrt{s}>$2.25 GeV. This result is 
rather different from the $\langle R^2 \rangle$ evaluation for the ELSA data 
presented in Fig.~\ref{ceta5}b). Besides the 2009 results \cite{Crede09}
(open circles) we consider here also the ELSA results from 2005~\cite{Crede05} (filled
squares). The latter measurement is afflicted with somewhat larger uncertainties
but it covers higher energies and, therefore, is interesting too for the present 
systematic analysis.
Note that a different scale is used in Fig.~\ref{ceta5}a), i.e. for the 
results corresponding to the CLAS data. 

After floating the normalization for the CLAS data from 2009 the agreement 
between experiment and the model predictions improves significantly. 
This is illustrated in Fig.~\ref{ceta5}c). Specifically, for $\sqrt{s}>$2.25 GeV
the values for ${<}R^2_{\rm mod}{>}$ (i.e. the $\chi^2$ per data point) are now, in general,
within the range of 1--2. In case of the ELSA data floating the normalization has
an influence too, but it is much less dramatic. Basically, the $\chi^2$ which 
amounted to values around 2 for energies above 2.25~GeV improves further to 
values around 1. 
Thus, once the normalization of the CLAS data is allowed to float there is a comparable
agreement of our model result with the CLAS 2009 and with the ELSA measurements. 
This suggests that all data available for the $\gamma{p}{\to}\eta p$ differential 
cross section are indeed mutually consistent, at least as far as their $t$-dependence 
is concerned. 

The renormalization factor $X$ is presented in Fig.~\ref{ceta5a}
as a function of the invariant collision energy. As can be seen, for the CLAS
data $X$ is not constant. It depends considerably on the energy in a rather peculiar
though still fairly smooth way. 
At the highest energy measured by the CLAS collaboration more than a factor two is
required to bring their data in agreement with the measurements at higher energies. 
In case of the ELSA 2009 data 
the normalization factor shows deviations in the order of 15\% from the nominal
value 1. Interestingly, one can see several strong statistical fluctuations. 
For the ELSA 2005 data the normalization factor differs even less from 1. 
In fact, for the higher energies $X$ is practically identical to the nominal value. 

More details about this statistical analysis are summarized in an Appendix. 
Specifically, there we provide numerical values for the achieved $\chi^2$ and
the normalization factor $X$ at each energy of the CLAS 2009 and the ELSA
2005 and 2009 measurements. 

Let us come back once again to the CLAS 2009 data. 
The dashed curves in Figs.~\ref{ceta4}-\ref{ceta2} indicate results where 
the renormalization factor is taken into account. Now there is a 
reasonable overall agreement between the experiment~\cite{Williams09} 
and the model calculation. It is worth noting that there are some 
remaining deviations at two ranges of the invariant collision energy. 
At 2.09$\le\sqrt{s}\le$2.13 GeV the data clearly show a
$t$-independent behaviour at $|t|$ above 1.3 GeV$^2$, say, which is
illustrated by the dash-dotted lines. This observation is in line with the
finding from the analysis of the ELSA data and might reflect the excitation 
of the $S_{11}(2090)$ resonance. Furthermore, at the
energy range 2.31$\le\sqrt{s}\le$2.54 GeV the experimental results show
some structure at $|t|\ge$ 1.3 GeV$^2$. Hints for such a structure are also
seen in the ELSA data from 2009, cf. Figs.~\ref{eta6c} and \ref{eta6b}. 
This could be a reflection of an excitation of a high spin resonance. 

In the context of the issues discussed in this section let us mention
a recent analysis of the $\gamma p\to\eta p$ reaction given in Ref.~\cite{He10}. 
The conclusion
of that study was that the recently published data by CLAS and ELSA in the
energy range 1.6$\le\sqrt{s}\le$2.8 GeV are well reproduced
due to the inclusion of Reggeized trajectories instead of simple $\rho$ and
$\omega$ poles. Obviously, our systematic analysis of $\eta$-meson 
photoproduction results does not support this statements from Ref.~\cite{He10} 
but reaches a different conclusion. 

\section{Summary}
We performed a systematic analysis of data available for 
differential cross sections of the reaction $\gamma{p}{\to}\eta p$.
In particular, we addressed the issue of 
consistency between the most recent data published by the 
ELSA~\cite{Crede09} and CLAS~\cite{Williams09} collaborations. 
Since the two measurements produced data points that 
are, in general, at different energies and four-momentum transfer 
squared or angles it is difficult to compare them directly.
Therefore, we utilized here as a link results of our Regge model 
that was fitted previously~\cite{Sibirtsev11} to the $\eta$-photoproduction 
data (differential cross sections and polarizations) at photon energies 
above 3 GeV ($\sqrt{s}>$2.55
GeV)~\cite{Anderson70,Braunschweig70,Dewire71,Bussey76,Bussey81}. 
Note that the latter aspect implies that our results may be considered 
as predictions with regard 
to the ELSA~\cite{Crede09} and CLAS~\cite{Williams09} measurements.

It was found out that our model reproduces the ELSA data from 2005~\cite{Crede05},
from 2008~\cite{Jaegle08}, and from 2009~\cite{Crede09} as well as
the CLAS results published in 2002~\cite{Dugger02} 
fairly well down to $\sqrt{s}\simeq$2.15~GeV.
At the same time we detected a substantial discrepancy
between the model calculations and the new CLAS data from 2009~\cite{Williams09}.
When floating the normalization of the new CLAS data it was possible to 
obtain a quite reasonable description of those data and bring them in line 
with our analysis of the other measurements. 
Indeed such a renormalization leads to a mutual consistency of all the 
$\gamma p\to\eta p$ data available presently for the differential cross
sections at invariant collision energies above approximately 2 GeV.
It turns out, however, that the renormalization factor depends on the energy.
Specifically, it increases more or less smoothly with energy and amounts to more 
than a factor two at the highest energy measured by the CLAS collaboration in 
2009. 

As a by-product we found promising indications for the presence of the 
$S_{11}(2090)$
excited baryon in form of an almost constant differential cross section at
$|t|>$1.3 GeV$^2$ within the range 2.03$\le\sqrt{s}\le$2.14 GeV. 
This observation is based on both the ELSA and CLAS measurements. We also 
detected hints for a resonance-like structure at the energy range 
2.31$\le\sqrt{s}\le$2.54 GeV covered by the CLAS data~\cite{Williams09}.

\begin{acknowledgement}
This work is partially supported by the Helmholtz Association through funds
provided to the virtual institute ``Spin and strong QCD'' (VH-VI-231), by
the EU Integrated Infrastructure Initiative  HadronPhysics2 Project (WP4
QCDnet) and by DFG (SFB/TR 16, ``Subnuclear Structure of Matter''). This
work was also supported in part by U.S. DOE Contract No. DE-AC05-06OR23177,
under which Jefferson Science Associates, LLC, operates Jefferson Lab. A.S.
acknowledges support by the JLab grant SURA-06-C0452  and the COSY FFE
grant No. 41760632 (COSY-085). 
\end{acknowledgement}

\section{Appendix}

In this Appendix we provide quantitative details of the comparison of our model 
results with
the data from the CLAS 2009 and the ELSA 2005 and 2009 measurements. Specifically, 
we list the $\chi^2$, for the original data points as well as after introducing a 
free normalization $X$ at each of the measured energies. 
The definition of 
the normalization constant $X$ and of the $\chi^2$ is given in Eq.~(\ref{chi2m}). 
The $\chi^2$ is obtained by considering all available data points in 
the range $|t|\le$2 GeV$^2$. 
We performed also fits to a smaller $t$ range, namely for $|t|\le$1 GeV$^2$,
guided by the idea that our Regge model might be even more reliable for such
small $t$ values. However, since often just 2 or 3 data points lie within
that range we refrain from giving a corresponding $\chi^2$ here. Rather
we use the $X$ obtained from the smaller $t$ range as an error estimate. 
The difference between the $X$ values obtained for the two $t$ ranges is
quoted as uncertainty in Tables~\ref{tab1} - \ref{tab3}.

Results for the CLAS 2009 data are given in Table~4. Comparing
columns 3 and 5 one can see the dramatic improvement in the
description once a normalization factor is introduced. Furthermore,
the difference between the CLAS data and the high-energy data,
as represented by our Regge model is large. It amounts to a 
factor of 2 and more for invariant collision energies above
2.50 GeV. When floating the normalization the $\chi^2$ reduces
to values around 1 to 2, which indicates that the $t$ dependence
found in the CLAS 2009 experiment is well in line with the one
exhibited by the high-energy data (and the Regge model). 
The $\chi^2$ is somewhat larger between 2.31 and 2.5 GeV where there
are indications for some structure in the $t$ dependence of the 
CLAS data that is not in line with the Regge prediction, cf. the
discussion in Sect. 5. 
For invariant collision energies below 2.15 GeV,
say, the $\chi^2$ improves only moderately when floating the
normalization and remains fairly large indicating the fact that our
Regge model is no longer able of describing the $t$ dependence
of the data. As expected, such a shortcoming cannot be compensated
by floating the normalization factor $X$. Moreover, then the value for
$X$ depends strongly on whether we fit the data up to $t$=-1 GeV$^2$
or -2 GeV$^2$, cf. the corresponding uncertainty given in the table. 

In case of the ELSA 2009 data there is also an improvement
in the $\chi^2$ when we float the normalization, cf. Table~\ref{tab2}. 
However, the
change is by far not as dramatic as for the CLAS 2009 data. Indeed,
in general, the normalization factor turns out to be within 
10 to 20\% only.
Also here we see an increasing deviation of the model predictions from
the $t$ dependence of the data for energies below 2.15 GeV. 

The ELSA 2005 measurement provided data for somewhat higher energies than
the one from 2009.
Interestingly, those data are in perfect agreement with the normalization
of our Regge model, fixed from older data for $\gamma p \to \eta p$ at
higher energies. The normalization factor found for the range 
2.30 - 2.50~GeV is practically identical to 1, cf. Table~\ref{tab3}. 
Again we see an increasing deviation of the model predictions from
the $t$ dependence of the data for energies below 2.15 GeV.

\begin{table}[t]
\begin{center}
\caption{\label{tab1}
Comparison of the CLAS data from 2009~\cite{Williams09}
with the results of our Regge model
for $|t|\le$2 GeV$^2$ at different photon energies $E_\gamma$. 
Values for the $\chi^2/N$ based on the published data are 
given in the third column. The renormalization factor $X$
obtained in a modified $\chi^2$ fit, cf. Eq.~(\ref{chi2m}), is given
in the fourth column while the corresponding $\chi^2/N$ can be found 
in the fifth column. The uncertainty of $X$ is explained in the text. 
}
\renewcommand{\arraystretch}{1.2}
\begin{tabular}{|c|c||c||c|c|}
\hline
$E_\gamma$ (GeV) & $\sqrt{s}$ (GeV) &  $\chi^2/N$
& $X$ & $\chi^2_{mod}/N$ \\
\hline
 3.695  &    2.79  &  45.0  &    0.40$\pm$0.03  & 1.0  \\
 3.431  &    2.70  &  55.4  &    0.39$\pm$0.03  & 2.2  \\
 3.302  &    2.66  &  41.0  &    0.42$\pm$0.04  & 1.8  \\
 3.190  &    2.62  &  27.3  &    0.47$\pm$0.03  & 0.9  \\
 3.079  &    2.58  &  26.1  &    0.50$\pm$0.04  & 0.9  \\
 2.970  &    2.54  &  27.0  &    0.50$\pm$0.02  & 1.4  \\
 2.862  &    2.50  &  32.4  &    0.50$\pm$0.01  & 1.9  \\
 2.757  &    2.46  &  20.9  &    0.57$\pm$0.02  & 1.4  \\
 2.653  &    2.42  &  19.4  &    0.59$\pm$0.03  & 2.4  \\
 2.550  &    2.38  &  16.8  &    0.63$\pm$0.06  & 2.2  \\
 2.475  &    2.35  &  16.6  &    0.63$\pm$0.09  & 2.8  \\
 2.425  &    2.33  &  16.7  &    0.62$\pm$0.04  & 1.4  \\
 2.375  &    2.31  &  18.5  &    0.61$\pm$0.06  & 1.6  \\
 2.326  &    2.29  &  15.9  &    0.61$\pm$0.04  & 0.8  \\
 2.278  &    2.27  &  17.2  &    0.63$\pm$0.02  & 1.5  \\
 2.229  &    2.25  &  23.5  &    0.60$\pm$0.03  & 2.2  \\
 2.182  &    2.23  &  25.2  &    0.58$\pm$0.03  & 1.6  \\
 2.134  &    2.21  &  25.1  &    0.59$\pm$0.00  & 1.6  \\
 2.087  &    2.19  &  23.6  &    0.60$\pm$0.03  & 1.7  \\
 2.041  &    2.17  &  27.0  &    0.60$\pm$0.05  & 2.2  \\
 1.995  &    2.15  &  27.5  &    0.61$\pm$0.07  & 3.7  \\
 1.949  &    2.13  &  24.7  &    0.64$\pm$0.10  & 4.8  \\
 1.909  &    2.11  &  26.9  &    0.71$\pm$0.17  &10.0  \\
 1.871  &    2.10  &  24.1  &    0.75$\pm$0.21  &10.3  \\
 1.848  &    2.08  &  28.2  &    0.78$\pm$0.25  &14.0  \\
 1.826  &    2.07  &  33.2  &    0.74$\pm$0.26  &14.2  \\
 1.804  &    2.06  &  41.6  &    0.72$\pm$0.25  &16.6  \\
 1.782  &    2.05  &  50.2  &    0.81$\pm$0.36  &24.0  \\
 1.760  &    2.04  &  30.2  &    0.80$\pm$0.27  &15.2  \\
 1.738  &    2.03  &  25.2  &    0.83$\pm$0.29  &13.2  \\
 1.717  &    2.03  &  33.0  &    0.81$\pm$0.30  &16.9  \\
 1.695  &    2.01  &   9.7  &    0.99$\pm$0.27  & 7.8  \\
 1.674  &    2.01  &  22.0  &    0.79$\pm$0.20  &10.5  \\
 1.653  &    2.00  &  17.6  &    0.79$\pm$0.17  & 8.2  \\
 1.631  &    1.98  &  16.1  &    0.79$\pm$0.15  & 7.1  \\
 1.610  &    1.97  &  14.8  &    0.76$\pm$0.12  & 5.5  \\
 1.589  &    1.96  &  15.1  &    0.77$\pm$0.12  & 5.9  \\
\hline
\end{tabular}
\end{center}
\end{table}

\begin{table}[t]
\begin{center}
\caption{\label{tab2}
Comparison of the ELSA data from 2009~\cite{Crede09}
with the results of our Regge model
for $|t|\le$2 GeV$^2$ at different photon energies $E_\gamma$. 
Values for the $\chi^2/N$ based on the published data are 
given in the third column. The renormalization factor $X$
obtained in a modified $\chi^2$ fit, cf. Eq.~(\ref{chi2m}), is given
in the fourth column while the corresponding $\chi^2/N$ can be found 
in the fifth column. The uncertainty of $X$ is explained in the text.
}
\renewcommand{\arraystretch}{1.2}
\begin{tabular}{|c|c||c||c|c|}
\hline
$E_\gamma$ (GeV) & $\sqrt{s}$ (GeV) &  $\chi^2/N$
& $X$ & $\chi^2_{mod}/N$ \\
\hline
 2.525  &    2.37  &   2.4  &    0.79$\pm$0.06  & 0.9  \\
 2.475  &    2.35  &   0.5  &    0.90$\pm$0.03  & 0.2  \\
 2.425  &    2.33  &   2.0  &    0.81$\pm$0.00  & 0.6  \\
 2.375  &    2.31  &   4.7  &    0.73$\pm$0.01  & 1.2  \\
 2.325  &    2.29  &   2.0  &    0.82$\pm$0.00  & 0.8  \\
 2.275  &    2.27  &   0.9  &    0.88$\pm$0.00  & 0.3  \\
 2.225  &    2.25  &   2.3  &    0.81$\pm$0.03  & 0.9  \\
 2.175  &    2.23  &   0.5  &    0.94$\pm$0.03  & 0.4  \\
 2.125  &    2.21  &   2.3  &    0.84$\pm$0.01  & 1.1  \\
 2.075  &    2.18  &   2.0  &    0.85$\pm$0.00  & 0.9  \\
 2.025  &    2.16  &   3.2  &    0.85$\pm$0.01  & 1.8  \\
 1.975  &    2.14  &   4.8  &    0.85$\pm$0.06  & 2.9  \\
 1.875  &    2.10  &  11.4  &    0.83$\pm$0.13  & 7.1  \\
 1.825  &    2.07  &  12.5  &    0.90$\pm$0.20  & 8.7  \\
 1.775  &    2.05  &  15.2  &    0.93$\pm$0.24  &10.8  \\
 1.725  &    2.03  &  17.7  &    0.94$\pm$0.26  &12.3  \\
 1.675  &    2.01  &  14.0  &    0.93$\pm$0.21  & 9.7  \\
 1.625  &    1.98  &   9.9  &    0.90$\pm$0.14  & 6.7  \\
 1.575  &    1.96  &   9.2  &    0.87$\pm$0.10  & 5.8  \\
\hline
\end{tabular}
\end{center}
\end{table}

\begin{table}[t]
\begin{center}
\caption{\label{tab3} Comparison of the ELSA data from 2005~\cite{Crede05}
with the results of our Regge model
for $|t|\le$2 GeV$^2$ at different photon energies $E_\gamma$. 
Values for the $\chi^2/N$ based on the published data are 
given in the third column. The renormalization factor $X$
obtained in a modified $\chi^2$ fit, cf. Eq.~(\ref{chi2m}), is given
in the fourth column while the corresponding $\chi^2/N$ can be found 
in the fifth column. The uncertainty of $X$ is explained in the text.
}
\renewcommand{\arraystretch}{1.2}
\begin{tabular}{|c|c||c||c|c|}
\hline
$E_\gamma$ (GeV) & $\sqrt{s}$ (GeV) &  $\chi^2/N$
& $X$ & $\chi^2_{mod}/N$ \\
\hline
 2.900  &    2.51  &   0.4  &    0.99$\pm$0.00  & 0.4  \\
 2.700  &    2.44  &   0.7  &    0.99$\pm$0.00  & 0.7  \\
 2.550  &    2.38  &   0.4  &    0.99$\pm$0.00  & 0.4  \\
 2.450  &    2.34  &   0.6  &    0.99$\pm$0.00  & 0.6  \\
 2.350  &    2.30  &   0.6  &    0.97$\pm$0.02  & 0.6  \\
 2.275  &    2.27  &   0.8  &    0.86$\pm$0.01  & 0.5  \\
 2.225  &    2.25  &   0.7  &    0.99$\pm$0.00  & 0.6  \\
 2.175  &    2.23  &   1.0  &    0.99$\pm$0.00  & 0.9  \\
 2.125  &    2.21  &   0.8  &    0.99$\pm$0.00  & 0.8  \\
 2.075  &    2.18  &   1.9  &    0.83$\pm$0.03  & 1.3  \\
 2.025  &    2.16  &   1.2  &    0.85$\pm$0.05  & 0.8  \\
 1.975  &    2.14  &   2.0  &    0.83$\pm$0.01  & 1.3  \\
 1.925  &    2.12  &   2.1  &    0.95$\pm$0.14  & 1.9  \\
 1.775  &    2.05  &   7.1  &    0.88$\pm$0.23  & 5.6  \\
 1.875  &    2.10  &   3.1  &    0.86$\pm$0.11  & 2.3  \\
 1.825  &    2.07  &   4.8  &    0.88$\pm$0.18  & 3.9  \\
 1.725  &    2.03  &   7.6  &    0.88$\pm$0.24  & 5.9  \\
 1.675  &    2.01  &   7.6  &    0.88$\pm$0.21  & 5.9  \\
 1.625  &    1.98  &   4.1  &    0.88$\pm$0.17  & 3.1  \\
 1.575  &    1.96  &   3.0  &    0.89$\pm$0.14  & 2.3  \\
\hline
\end{tabular}
\end{center}
\end{table}


\end{document}